\documentclass[11pt]{article}

\usepackage[final]{acl}

\usepackage{times}
\usepackage{latexsym}
\usepackage{microtype}
\usepackage{enumitem}
\usepackage{subcaption}
\usepackage{bbding}
\usepackage{pifont}
\usepackage{lineno}
\usepackage{multirow}
\usepackage[export]{adjustbox}
\usepackage{colortbl}
\usepackage{cuted,tcolorbox}
\usepackage{multicol}
\usepackage{wrapfig}
\usepackage{soul}
\usepackage[skip=1ex]{caption}

\usepackage{amsmath,amsfonts,bm}

\def\eqref#1{equation~\ref{#1}}

\def\1{\bm{1}}

\def\rmA{{\mathbf{A}}}

\def\rmL{{\mathbf{L}}}
\def\rmM{{\mathbf{M}}}

\def\rmX{{\mathbf{X}}}

\def\vtheta{{\bm{\theta}}}

\DeclareMathAlphabet{\mathsfit}{\encodingdefault}{\sfdefault}{m}{sl}
\SetMathAlphabet{\mathsfit}{bold}{\encodingdefault}{\sfdefault}{bx}{n}

\usepackage[utf8]{inputenc} 
\usepackage[T1]{fontenc}    
\usepackage{hyperref}       
\usepackage{url}            
\usepackage{booktabs}       
\usepackage{amsfonts}       
\usepackage{nicefrac}       
\usepackage{microtype}      
\usepackage{xcolor}         
\usepackage{geometry}
\usepackage{listings}
\usepackage[ruled,vlined]{algorithm2e}

\newcommand{\highlight}[1]{\textcolor{red}{#1}}

\newtheorem{definition}{Definition}[section]

\newcommand{\sysname}{\textsc{UserMirrorer}}

\definecolor{myblue}{HTML}{4285F4}
\definecolor{myred}{HTML}{EA4335}

\usepackage[T1]{fontenc}

\usepackage[utf8]{inputenc}

\usepackage{microtype}

\usepackage{inconsolata}

\usepackage{graphicx}

\title{Mirroring Users: Towards Building Preference-aligned User Simulator with User Feedback in Recommendation}

\author{
  \textbf{Tianjun Wei}$^{1}$\quad
  \textbf{Huizhong Guo}$^{2}$\quad
  \textbf{Yingpeng Du}$^{1}$\thanks{Corresponding author.}\quad
  \textbf{Zhu Sun}$^{3}$\\
  \textbf{Huang Chen}$^{1}$\quad
  \textbf{Dongxia Wang}$^{2}$\quad
  \textbf{Jie Zhang}$^{1}$\\
  $^{1}$Nanyang Technological University, Singapore \\
  $^{2}$Zhejiang University, Hangzhou, China \\
  $^{3}$Singapore University of Technology and Design, Singapore \\
  \texttt{tjwei2-c@my.cityu.edu.hk},~\texttt{huiz\_g@zju.edu.cn},~\texttt{dyp1993@pku.edu.cn},\\
  \texttt{zhu\_sun@sutd.edu.sg},~\texttt{huang012@e.ntu.edu.sg},\\\texttt{dxwang@zju.edu.cn},~\texttt{zhangj@ntu.edu.sg}\\
}

\begin{document}
\maketitle
\begin{abstract}
User simulation is increasingly vital to develop and evaluate recommender systems (RSs). While Large Language Models (LLMs) offer promising avenues to simulate user behavior, they often struggle with the absence of specific task alignment required for RSs and the efficiency demands of large-scale simulation. A vast yet underutilized resource for enhancing this alignment is the extensive user feedback inherent in RSs, but leveraging it is challenging due to its ambiguity, noise and massive volume, which hinders efficient preference alignment. To overcome these hurdles, we introduce a novel data construction framework that leverages user feedback in RSs with advanced LLM capabilities to generate high-quality simulation data. Our framework unfolds in two key phases: (1) using LLMs to generate decision-making processes as explanatory rationales on simulation samples, thereby reducing ambiguity; and (2) data distillation based on uncertainty estimation and behavior sampling to efficiently filter the most informative, denoised samples. Accordingly, we fine-tune lightweight LLMs, as user simulators, using such high-quality dataset with corresponding decision-making processes. Extensive experiments confirm that our framework significantly boosts the alignment with human preferences and the in-domain reasoning capabilities of the fine-tuned LLMs, providing more insightful and interpretable signals for RS interaction. We believe our work, together with publicly available developed framework, high-quality mixed-domain dataset, and fine-tuned LLM checkpoints, will advance the RS community and offer valuable insights for broader human-centric AI research. Our code is available at \url{https://github.com/Joinn99/UserMirrorer}.
\end{abstract}

\section{Introduction}
User-centric AI, which uses behavior analysis to understand human preferences, is fundamental to research in recommender systems (RSs), social networks, and search engines~\citep{wangUserBehavior2025, adomaviciusNextGeneration2005, gaoS3Socialnetwork2023, wangIndepthInvestigation2024}. As a crucial branch, RSs are vital in web services, mitigating information overload by continuously capturing user preferences from interaction histories (e.g., products, content)~\citep{wangSurveyFairness2023,WuEfficientLarge2026}.
RS development, as a direct AI service, critically depends on user feedback. While online testing remains the gold standard for real user feedback capture~\citep{dengTrustworthyAnalysis2017}, its practical limitations are significant: (1) Achieving statistical significance requires prolonged data collection (weeks/months), substantially delaying algorithm iteration~\citep{gilotteOfflineTesting2018}; (2) Increasing privacy concerns challenge traditional user data-dependent evaluations~\citep{stavinovaSyntheticDataBased2022a}.

These constraints call for virtual replicas of users that can faithfully reproduce real behavior without direct human involvement. The concept of digital twins, which creates virtual counterparts of real-world entities for simulation and optimization, provides a principled framework for this goal~\citep{aliCognitiveDigitalTwins2021}.
Originally rooted in industrial systems such as smart manufacturing~\citep{aliCognitiveDigitalTwins2021} and battery analytics~\citep{zhuPhysicsInformedMultiTask2026}, digital twins have evolved toward cognitive modeling that replicates not only observable states but also the reasoning and decision-making of complex agents.
In the context of RSs, such cognitive digital twins take the form of user simulators~\citep{ieRecSimConfigurable2019}, mechanisms that generate synthetic yet faithful user behavior data to enable efficient, privacy-preserving evaluation and optimization.

User simulators generate synthetic profiles to simulate user-system interactions, enabling the offline evaluation of RSs. Earlier methods primarily used rule-based \citep{weiAirDialogueEnvironment2018} or reinforcement learning approaches \citep{ieRecSimConfigurable2019,zhaoKuaiSimComprehensive2023}.
However, these traditional simulators are limited: they focus only on intra-system preferences, neglecting essential external contextual knowledge and the reasoning underlying real user decisions.
Recent advances in Large Language Models (LLMs) \citep{teamGemma22024}, with their extensive world knowledge and advanced multi-step reasoning capabilities \citep{deepseek-aiDeepSeekR1Incentivizing2025,guoThinkWhen2026}, offer a solution. They have enabled the development of simulators with human-like cognitive models in various fields, including  paper review~\citep{jinAgentReviewExploring2024}, social network~\citep{gaoS3Socialnetwork2023}, and even smart manufacturing~\citep{aliCognitiveDigitalTwins2021}.
Research on LLM-based user simulators generally follows two paths: (1) using LLMs' generalization for feature extraction in traditional simulators \citep{coreccoSUBERRL2024, zhangLLMPoweredUser2025, LiuDiagnosticGuided2026}, and (2) creating autonomous agents to simulate fine-grained individual user behaviors in recommendation scenarios \citep{zhangGenerativeAgents2024, zhangUSimAgentLarge2024, wangRecMindLarge2024, zhangAgentCFCollaborative2024}.
Despite their promise, these LLM-based approaches have two major drawbacks:
First, they over-rely on pre-trained LLM knowledge without fine-tuning on massive user feedback, resulting in poor task adaptation.
Second, the need for a powerful LLM backend introduces prohibitive computational costs for large-scale user simulations in RSs due to the lack of task-specific knowledge.

In this work, we explore how to harness user feedback from RSs to fine-tune LLM-based user simulators, thereby improving their fidelity to authentic user behavior. 
In the context of RSs, raw user feedback exhibits two distinctive characteristics that present unique challenges on the development of LLM-based user simulators:
\begin{itemize}[leftmargin=12pt]
\item User feedback typically captures only observed behaviors, omitting the underlying \textbf{decision-making processes}. This absence of context introduces \textbf{ambiguity} and hinders the detection of \textbf{noisy} data, limiting LLM from learning the human decision-making process.
\item RSs generate \textbf{massive volumes of feedback with highly variable quality}, making it essential to filter and select \textbf{high-quality} samples to achieve effective and efficient LLM fine-tuning.
\end{itemize}
To address these challenges, we propose an end-to-end framework that fully exploits user feedback in RSs to align LLM simulators with genuine user preferences, named \sysname.
Specifically, our framework first transforms feedback from diverse domains into a unified simulation scenario, capturing user memory, item exposures, and associated behaviors.
Then, by prompting LLMs to generate explicit decision-making processes as explanatory rationales, we can reduce feedback ambiguity, denoise behavior signals, and quantify LLM performance across contexts.
Finally, by selecting scenarios that expose the limitations of weaker LLMs, and matching decision processes with real user behaviors, we can curate a refined dataset comprising \textbf{challenging 
samples, high-fidelity decision processes generated by powerful LLMs, and denoised behavioral signals.}

Based on our developed framework, we investigate the following research questions:
(RQ1) \textit{Does the constructed dataset enable LLM simulator to make actions that align with human preferences with explanatory rationales?}
(RQ2) \textit{Does the fine-tuned user simulator more effectively mimic user behavior and provide enhanced support to the RSs?}
To answer these questions, we construct a high-quality dataset from eight diverse domains and fine-tune LLMs to derive user simulators.
Experimental results 
demonstrate that these LLM simulators achieve significant improvements in alignment with real user preferences, and able to provide effective and interpretable feedback to RSs. 

We are committed to open-sourcing our framework\footnote{\url{https://github.com/Joinn99/UserMirrorer}}, the constructed dataset\footnote{\url{https://huggingface.co/datasets/Joinn/UserMirrorer}}, and the fine-tuned LLM-based user simulators\footnote{\url{https://huggingface.co/Joinn/UserMirrorrer-Llama-DPO}}, aiming to provide the academic and industrial communities with a tool that effectively enhances the realism and efficiency of user simulators.
Our framework enables developers to build customized domain-specific user simulators, thereby advancing research and applications in RSs and beyond.

\section{Building User Simulation Scene from User Feedback}
\subsection{Task formulation}
In real-world internet applications, user behaviors are highly diverse, extending beyond simple clicks to include complex interactions such as ratings, writing reviews, and sharing content. Considering the core focus of RSs and the primary objective of user simulators, we concentrate on a prevalent and generalizable impression-aware scenario~\citep{yangTopkRecommendation2012}.
In this scenario, a Recommender System (RS) presents a user with an exposure (a list of items). The user then acts on this exposure based on their preferences (e.g., clicking an item). While some existing works independently predict the user's interaction probability for each item \citep{zhangLLMPoweredUser2025,zhangGenerativeAgents2024}, we adopt a listwise approach \citep{zhaoKuaiSimComprehensive2023,dengOneRecUnifying2025a}. This means we simulate user behavior at the level of the entire exposure, more closely mirroring how humans interact in reality.
To emulate this behavior, the user simulator must rely on the user's current state. This state comprises their profile, historical interactions, and potential emotional factors that influence decisions \citep{zhangLLMPoweredUser2025, wangUserBehavior2025}. We collectively term this comprehensive information the user simulator's \textbf{"memory"}. Based on the above definitions, the user simulation scene we focus on can be formally defined as follows:
\begin{definition}[User Simulation Scene in RSs]
Let $\rmM$ denote the user’s memory (e.g., profile and interaction history) and let $\rmL = \{\,l_1,\dots,l_k\}$ 
be the exposure list presented to the user. We form a simulation scene $\rmX = \mathrm{Prompt}(\rmM, \rmL)$. The action space is $\rmA = \{\,a_1,\dots,a_k\}$, where action $a_i$ corresponds to interacting with item $l_i$. A user simulator parameterized by $\theta$ defines a categorical distribution over $\rmA$ conditioned on $\rmX$:
\begin{equation}
\footnotesize
P_\theta(a_i \mid \rmX)\in[0,1],
\qquad
\sum\nolimits_{i=1}^k P_\theta(a_i\mid \rmX)=1.
\end{equation}
Equivalently, we write
$P_\theta(\rmA\mid \rmX)=\bigl(P_\theta(a_1\mid \rmX),\dots,P_\theta(a_k\mid \rmX)\bigr)$.
The goal of the simulator is to learn $P_\theta(\cdot\mid \rmX)$ so as to match the distribution of real user behaviors.
\end{definition}
Existing LLM-based simulators typically focus on augmenting the user's contexts by utilizing the knowledge and reasoning capabilities of the LLM as well as the agent's memory mechanisms~\citep{wangRecMindLarge2024, zhangAgentCFCollaborative2024, zhangGenerativeAgents2024, zhangUSimAgentLarge2024}. Comparatively, this research emphasizes the exploration of fine-tuning the LLM to \textbf{better match human preferences in certain user simulation scene}. The fine-tuned LLM can be used as a plugin to integrate with the existing LLM agent-based user simulator work to achieve better alignment with real user preference and higher efficiency.

\subsection{Creating user simulation scenes}

\begin{figure}[t]
\centering
    \includegraphics[width=\linewidth]{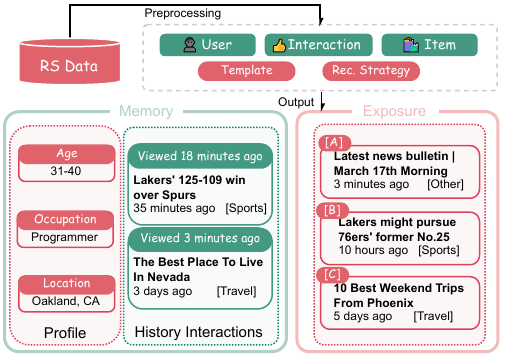}
    \caption{Example of the user simulation scene.}
    \label{Fig:Scene}
\end{figure}

To effectively leverage the massive user feedback data for training and aligning the LLM user simulator, we follow the ideas from existing works based on LLM agents~\citep{coreccoSUBERRL2024,zhangGenerativeAgents2024}. We organize the raw feedback data into an user simulation scene that an LLM agent can understand. The core elements of a user simulation scene include two key components: \textbf{memory} and \textbf{exposure}, as shown in Figure \ref{Fig:Scene}.

\textbf{Memory}. Follow existing work~\citep{coreccoSUBERRL2024}, we consider two main components in memory: user profile and interaction history. The user profile contains the user's static attributes, and interaction history record a series of past interactions with items. To accommodate the varying information granularity across different datasets, we develop a series of domain-specific templates.
These templates systematically convert the available user attributes and interacted item features within each dataset into a standardized plain text format, for example, \textit{"Age: 35-44, Occupation: Customer Service, User Interaction History: Crimson Tide (1995) Rating: 4/5 ..."} (Please refer to Figure \ref{Fig:Scene}). This unified textual representation then serves as the memory input for the user simulator, enabling it to accurately model user preferences. Detailed templates for all datasets can be found in the Appendix \ref{subsec:appendix_prompts}.

\textbf{Exposure}. In most public RS datasets, typically only information about items with which the user actually interacted is available, lacking the complete exposure the user saw when making their decisions~\citep{perezmaureraImpressionAwareRecommender2025}. 
Therefore, for datasets missing this information, we utilize existing recommendation strategies to construct an exposure list.  
To ensure this list more accurately reflects items a user might genuinely prefer, we build a hybrid exposure list based on the user's interaction behavior. This list is generated by uniformly sampling $K$ items from three distinct sources: random sampling, collaborative filtering~\citep{shen2021powerful} and content-based nearest neighbor retrieval~\citep{deshpandeItembasedTopN2004}
The specific implementation details and parameter configurations are  described in Appendix \ref{subsec:appendix_exposure_list}.
Finally, we randomly sample $N$ items in total from all Top-$K$ item lists and randomly insert the ground truth item into an arbitrary position within this list, resulting in a final exposure list of length $N+1$. We set $K$=32 and $N$ is uniformly sampled in the range of 2 to 12 across different users to ensure the generalization capability. 

Upon completing the construction of the user memory and exposure, we establish a scene for the LLM user simulator. The user simulator then takes the corresponding behavior from the exposures. To aid our subsequent analysis of the LLM's behaviors, we adopt practices from existing ranking and search-related works~\citep{sunChatGPTGood2023,chenAttentionLarge2025} to use alphabetical identifiers such as \textit{[A], [B], [C]}, etc., to label each option in the exposures. This approach encourages the LLM to output clear option identifiers rather than free-form text descriptions.
In our developed framework \sysname, the scene construction process uses domain-specific templates to transform user feedback data from various domains into a unified format that LLMs can easily interpret.
Specifically, we design the templates for datasets from eight different domains, including movies, books, news, etc., to convert the data to simulation scenes for further experiment and analysis. Detailed descriptions, statistics and template used for all datasets can be found in the Appendix \ref{subsec:appendix_dataset} and \ref{subsec:appendix_prompts}.

\section{\sysname: Building a Preference-aligned User Simulator}
\subsection{Can user feedback be helpful in user preference alignment?}
\begin{figure}[t]
    \centering
    \includegraphics[width=0.95\linewidth]{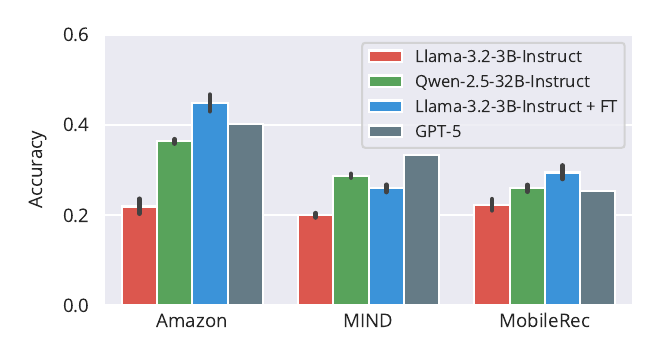}
    \caption{Accuracy of user simulation on representative domains with different base LLMs.}
    \label{Fig:Direct}
\end{figure}

To validate the effectiveness of leveraging user feedback data in aligning user simulators with real user preferences and to explore the factors that may influence this process, we conduct a straightforward preliminary experiment. Specifically, we randomly sample 1,024 user simulation scenes each for the training and test sets, and fix the exposure list length at 5.
We employ different base LLMs as user simulators, and fine-tune a \textit{Llama-3.2-3B-Instruct} model for 1 epoch using the construct training set (See Appendix \ref{subsec:finetune_config} for fine-tuning configurations).
We measure accuracy by matching LLM-predicted behaviors against actual user behaviors, and report representative results from three domains in Figure~\ref{Fig:Direct}.

\textbf{Stronger models perform better.} Experimental results consistently show that LLMs with more advanced foundational capabilities achieve higher alignment with real user behaviors.  Since these LLMs have not been fine-tuned on domain-specific user feedback in RSs, their more human-like behaviors likely stem from their robust context modeling and reasoning capabilities, enabling them to more effectively extract relevant information from the user’s memory in relation to the current exposure.
This indicates that the intrinsic capabilities of LLMs serve as a crucial foundation for their successful simulation of user preferences.


\textbf{Fine-tuning using real user feedback works.} The most significant finding is that directly utilizing user feedback data for fine-tuning is indeed effective. Figure \ref{Fig:Direct} shows that after fine-tuned, the accuracy of the user behavior prediction shows a significant improvement. This demonstrates that the real user feedback data may contain latent and consistent preferences across different users.

However, it is also observed that on the MIND dataset, the stronger \textit{Qwen-2.5-32B-Instruct} and \textit{GPT-5} still achieves better alignment with real user preferences than the fine-tuned weaker LLM.
This suggests that the current data alone might be insufficient for optimal preference alignment, and the performance still can be potentially improved.

\subsection{Data distillation based on uncertainty decomposition}
\label{Subsec:Data_Distillation}
As demonstrated above, harnessing the synergy between powerful LLMs’ intrinsic reasoning capabilities and the implicit preferences in user feedback can markedly align user simulator with user preference.
However, the nature of user feedback in RSs may pose following challenges:
\begin{itemize}[leftmargin=12pt]
    \item \textbf{First}, user “memory” from profiles and past interactions lacks key context (e.g., environment, mental state), so the user intent remains unclear, bringing \textbf{ambiguity} to the data. 
    \item \textbf{Second}, raw logs often include \textbf{noise} (e.g., accidental clicks) that brings negative effects. 
    \item \textbf{Third}, behaviors stem from hidden decision steps, from impulse to deliberation, require strong LLM reasoning abilities to handle. Hence, it will be difficult to directly leverage user feedback to \textbf{transfer such capabilities} from a powerful LLM to a lightweight LLM, to meet the efficiency requirements of user simulation in RSs.
\end{itemize}

Given the limitations inherent in user feedback and the vast amount of feedback data in real-world RSs, it may not be optimal or \textbf{efficient} to fine-tune an LLM user simulator using all available raw data. Instead, selecting high-quality data to address these limitations could be a more effective approach.
In this section, we seek to distill a curated subset of high-quality training samples, and propose a solution that kills two birds with one stone. 

\textbf{Motivation.} In~\citet{houDecomposingUncertainty2024}, the author suggests that through generating clarifications, we can effectively distinguish and quantify two main sources of uncertainty during LLM generating process~\citep{galDropoutBayesian2016}: \textit{aleatoric (data) uncertainty}, which arises from inherent data ambiguity, and \textit{epistemic (model) uncertainty}, which stems from LLM's limitations in knowledge and capability. Building on this,~\citet{houDecomposingUncertainty2024} suggest to generate multiple clarifications for inputs, thereby reducing the ambiguity of inputs with clarification and better estimate the model uncertainty. 
\begin{definition}[Uncertainty Decomposition]
Assume $P(\bm C | \bm X)$ is distribution of the generated clarifications given a input $\mathbf{X}$. Then, the prediction is defined as the ensemble of predictions conditional on input with clarifications, i.e., $P(\bm Y | \bm X)=\mathbb{E}_{P(\bm C | \bm X)}[P(\bm Y | \bm X \oplus \bm C, \vtheta)]$ (Model parameter $\vtheta$ is a constant and will be omitted for brevity below.), and the decomposed uncertainty can be represented as:
\begin{equation}
    \footnotesize
   \underbrace{\mathcal{H}(P(\bm Y | \bm X))}_{\text{\ding{172}}} =\underbrace{\mathcal{I}(\bm Y; \bm C | \bm X)}_{\text{\ding{173}}} + \underbrace{\mathbb{E}_{P(\bm C | \bm X)}\mathcal{H}(P(\bm Y | \bm X \oplus \bm C))}_{\text{\ding{174}}},
   \label{eq:our_decompose}
\end{equation}
where $P(\bm Y | \bm X \oplus \bm C)$ denote the distribution of output with the clarification $\bm C$, $\mathcal{H}$ denotes the entropy function, $\mathcal{I}(\bm Y; \bm C | \bm X)$ denotes the mutual information between the output distribution $\bm Y$ and clarifications $\bm C$. Here, \ding{173} is the approximation of \textbf{aleatoric uncertainty}, \ding{174} represents the \textbf{epistemic uncertainty}, and \ding{172} denotes the total uncertainty. 
In practice, the expectation terms are approximated via the ensemble of $N$ clarifications sampled from the same input $\rmX$.
\label{Def:Uncert}
\end{definition}

\begin{figure}[t]
    \centering
    \includegraphics[width=\linewidth]{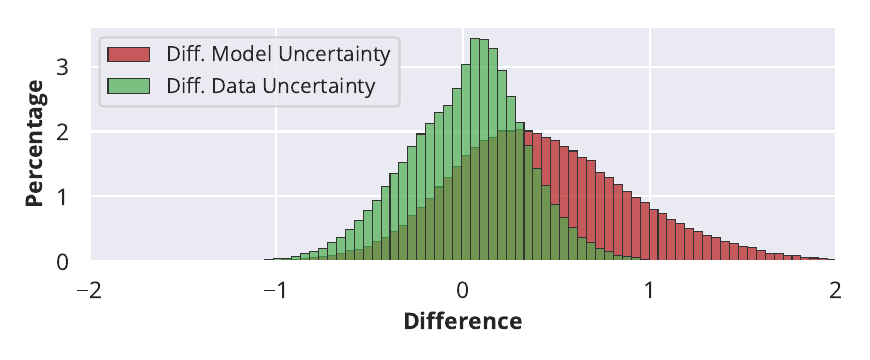}
    \caption{Results of uncertain decomposition on user simulation scenes.}
    \label{Fig:UncertainDecomp}
\end{figure}

\begin{figure*}[t]
    \begin{center}
    \includegraphics[width=0.75\linewidth]{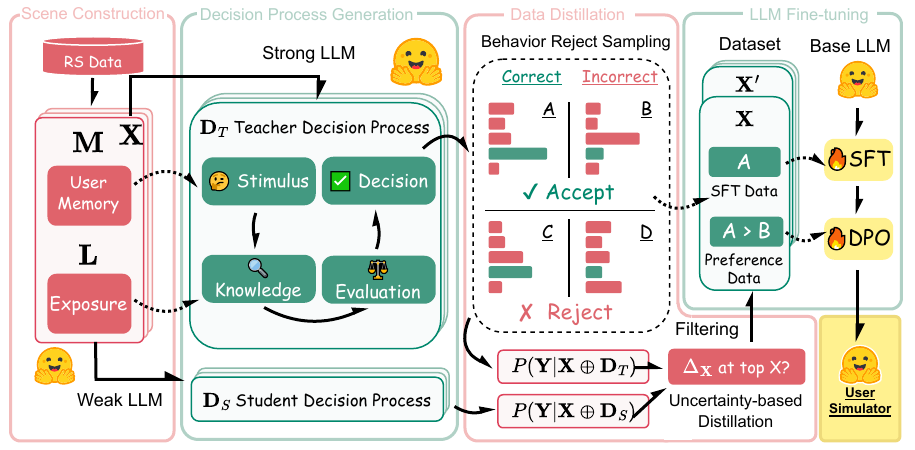}
    \caption{Illustration of our proposed \sysname~framework.}
    \label{Fig:Pipeline}
\end{center}
\end{figure*}

\textbf{Decision-process Generation.} Following this idea, we aim to decompose the uncertainty lied in user behavior simulation.
When a human user provides feedback, there is always an underlying decision-making process. We believe that a complete decision-making process can play a "clarifying" role in user simulation. We adopt the widely researched consumer behavior model, the \textbf{\textit{Engel Kollat Blackwell (EKB)}} model~\citep{engelConsumerBehavior1978}, with appropriate adaptations for user simulation scenes.
This model was originally used to explain the decision-making processes of user purchasing, encompassing the complete journey from need recognition to post-purchase evaluation. To make it more suitable to user simulation in RSs, we have slightly adjust the steps of the model. The complete decision-making process includes the following stages:
\begin{itemize}[leftmargin=12pt]
    \item \textbf{Stimulus}: Identify the factors triggering the user's interaction, including external (time, location, social context) and internal (needs, emotional state) elements.
    \item \textbf{Knowledge}: Extract relevant factors and attributes from the exposures for further analysis.
    \item \textbf{Evaluation}: Evaluate potential behaviors using different styles \citep{kahnemanThinkingFast2011}, from intuitive to logical.
\end{itemize}

After generating the above decision-making process, the LLM user simulator will predict the corresponding user behavior (See Appendix \ref{subsec:appendix_prompts} for full prompt and \ref{Append:Case} for example case). This decision process furnishes the user simulation scenario with essential contextual information, thereby \textbf{reducing ambiguity} in raw user feedbacks.

\textbf{Uncertainty-based Data Distillation.} Moreover, when the decision process effectively leverages information from the user’s memory, the simulator should choose its behavior with high confidence, reflecting the model’s epistemic uncertainty.
Therefore, by comparing the epistemic uncertainties of different LLMs on the same scene, we can \textbf{quantify their relative capabilities in specific user simulation scenes}, and select challenging cases where strong and weak LLMs exhibit larger difference in epistemic uncertainties.
Suppose the decision-making process of scene $\bm X$ generated by LLMs $A$ and $B$ as $\bm D_A$ and $\bm D_B$ respectively, the difference of epistemic uncertainties is defined as:
\begin{equation}
\footnotesize
\begin{aligned}
\Delta_{EU}(\bm X, (A, B)) = &\mathbb{E}_{P(\bm D_A | \bm X)}\mathcal{H}(P(\bm Y | \bm X \oplus \bm D_A)) \\-& \mathbb{E}_{P(\bm D_B | \bm X)}\mathcal{H}(P(\bm Y | \bm X \oplus \bm D_B)).
\end{aligned}
\label{Eq:DiffUncert}
\end{equation}

To validate effectiveness of the above method in uncertainty decomposition, we have conducted an experiments. We randomly sample 16k user simulation scenes from each dataset, and employ \textit{Llama-3.2-3B-Instruct} and \textit{Qwen2.5-32B-Instruct} as the weaker and stronger models, respectively, prompting each to generate $N=10$ decision-making processes per scene. (See Appendix~\ref{Append:BuiltDataset} for the details of configurations of LLM inference).
We then use the generated processes to guide \textit{Llama-3.2-3B-Instruct} in predicting user behavior and compute the uncertainty difference based on the log-probabilities of its output options.
Figure \ref{Fig:UncertainDecomp} shows that weaker LLMs tend to have higher epistemic uncertainty across most scenes. Scenes with large uncertainty gaps between stronger and weaker models are likely more complex, subtle, or involve nuanced decision-making—areas where weak LLMs often struggle. Therefore, we believe that focusing on these challenging cases will yield more effective datasets for improving the LLM user simulator.

\textbf{Sampling Denoised Behaviors.} Although the distillation process generates challenging simulation scenes, they may still contain noisy behaviors. Viewing each generated decision process as a rationale for its predicted action, we employ rejection sampling to remove potential noise. Specifically, for each scene, we compare the predictions regarding to $N$ decision-making processes to the actual user behavior. We consider the scenes for which none of the decision processes yields a correct match as the potential noisy behaviors, and discard it. 
For the remaining scene, we mark the decision process according to the matching of behaviors as \textit{accepted} and \textit{rejected}, and formulate the samples with the highest confidence as the preference pair.

The full dataset construction pipeline, is illustrated in Figure \ref{Fig:Pipeline}, and the detailed steps are shown in Appendix \ref{Append:DatasetPipeline}.

\section{Experiments}
In this section, we conduct comprehensive evaluations of our data construction pipeline and the fine-tuned LLM user simulator. Our analysis focuses on the two key research questions raised: 
\begin{itemize}[leftmargin=12pt]
    \item \textbf{RQ1}: \textit{Can our constructed dataset further align LLM user simulators with human preferences?} 
    \item \textbf{RQ2}: \textit{Can the fine-tuned user simulator provide effective and interpretable feedback for RSs?}
\end{itemize}

\begin{table*}[t]
    \centering
    \renewcommand{\arraystretch}{0.85}
    \scriptsize
    \setlength\tabcolsep{2pt}
    \setul{1pt}{0.4pt}
    \begin{tabular}{l|l|c|cccccccc}
        \toprule
         & \multicolumn{1}{c|}{\textbf{Type of Exposure}} & \textbf{Real Exposure} & \multicolumn{8}{c}{\textbf{Synthetic Exposure}} \\ \midrule
         & \multicolumn{1}{c|}{\textbf{Dataset}} & \textbf{MIND} & \textbf{Amazon} & \textbf{KuaiRec2} & \textbf{LastFM} & \textbf{Goodreads} & \textbf{Movielens} & \textbf{MobileRec} & \multicolumn{1}{c|}{\textbf{Steam}} & \textbf{Overall} \\ \midrule
         & \textbf{\:\:Gemma-3-4B-it} & 27.5 & 24.8 & 27.7 & 19.8 & 50.0 & 42.2 & 26.1 & \multicolumn{1}{c|}{31.1} & 29.6 \\
         & \textbf{\:\:Gemma-3-12B-it} & 29.1 & 34.4 & 26.8 & 26.8 & 56.5 & 55.0 & 23.1 & \multicolumn{1}{c|}{33.8} & 36.0 \\
         & \textbf{\:\:Qwen2.5-7B-Instruct} & 25.9 & 29.7 & 28.1 & 24.9 & 57.2 & 46.9 & 27.0 & \multicolumn{1}{c|}{33.9} & 33.7 \\
         & \textbf{\:\:Qwen2.5-14B-Instruct} & 27.9 & 31.5 & 32.3 & 24.5 & 63.1 & 55.3 & 26.0 & \multicolumn{1}{c|}{29.2} & 35.6 \\
        \multirow{-8}{*}{\textbf{\begin{tabular}[c]{@{}l@{}}Base\\ Models\end{tabular}}} & \textbf{\:\:Qwen2.5-32B-Instruct (Teacher)} & 28.7 & 36.3 & 31.1 & 27.3 & 70.7 & 54.6 & 26.0 & \multicolumn{1}{c|}{42.4} & 39.7 \\ \midrule
         & \textbf{\:\:Qwen2.5-3B-Instruct} & 27.2 & 23.7 & 26.4 & 22.8 & 41.1 & 33.2 & 22.4 & \multicolumn{1}{c|}{31.2} & 27.2 \\
         & \textbf{\:\:+SFT} & 30.9 & 43.3 & 38.2 & 32.6 & 76.2 & 57.4 & 30.4 & \multicolumn{1}{c|}{56.6} & 46.4 \\
         & \textbf{\:\:+SFT+GRPO} & \cellcolor[HTML]{D5E8D4}32.8 & \cellcolor[HTML]{D5E8D4}49.9 & 44.9 & \cellcolor[HTML]{FFF2CC}\ul{36.0} & \cellcolor[HTML]{D5E8D4}85.3 & 64.6 & \cellcolor[HTML]{D5E8D4}32.5 & \multicolumn{1}{c|}{67.7} & \cellcolor[HTML]{D5E8D4}53.1 \\
         & \textbf{\:\:+SFT+DPO} & \cellcolor[HTML]{FFF2CC}\ul{33.5} & \cellcolor[HTML]{F8CECC}\textbf{52.3} & \cellcolor[HTML]{D5E8D4}46.2 & 35.3 & \cellcolor[HTML]{FFF2CC}\ul{86.3} & \cellcolor[HTML]{FFF2CC}\ul{66.2} & \cellcolor[HTML]{D5E8D4}32.5 & \multicolumn{1}{c|}{\cellcolor[HTML]{F8CECC}\textbf{71.3}} & \cellcolor[HTML]{FFF2CC}\ul{54.7} \\ \cmidrule(l){2-11} 
         & \textbf{\:\:Llama-3.2-3B-Instruct} & 19.9 & 21.9 & 22.5 & 18.3 & 27.1 & 23.3 & 22.1 & \multicolumn{1}{c|}{26.2} & 22.7 \\
         & \textbf{\:\:+SFT} & 27.3 & 44.1 & 40.2 & 28.5 & 74.1 & 51.5 & 30.0 & \multicolumn{1}{c|}{59.8} & 46.0 \\
         & \textbf{\:\:+SFT+GRPO} & 32.2 & 48.4 & \cellcolor[HTML]{FFF2CC}\ul{46.6} & \cellcolor[HTML]{F8CECC}\textbf{38.2} & 82.7 & \cellcolor[HTML]{D5E8D4}65.3 & \cellcolor[HTML]{FFF2CC}\ul{32.6} & \multicolumn{1}{c|}{\cellcolor[HTML]{D5E8D4}68.0} & 52.7 \\
        \multirow{-12}{*}{\textbf{\begin{tabular}[c]{@{}l@{}}Fine-tuned\\ Models\end{tabular}}} & \textbf{\:\:+SFT+DPO} & \cellcolor[HTML]{F8CECC}\textbf{34.0} & \cellcolor[HTML]{FFF2CC}\ul{51.8} & \cellcolor[HTML]{F8CECC}\textbf{48.7} & \cellcolor[HTML]{D5E8D4}35.7 & \cellcolor[HTML]{F8CECC}\textbf{86.9} & \cellcolor[HTML]{F8CECC}\textbf{67.3} & \cellcolor[HTML]{F8CECC}\textbf{34.7} & \multicolumn{1}{c|}{\cellcolor[HTML]{FFF2CC}\ul{70.0}} & \cellcolor[HTML]{F8CECC}\textbf{55.0} \\ \midrule
        & \textbf{GPT-5-Nano (2025-08-07)} & 28.4 & 35.4 & 31.9 & 27.8 & 65.6 & 57.8 & 24.5 & \multicolumn{1}{c|}{33.3} & 36.8 \\
        & \textbf{GPT-5-Mini (2025-08-07)} & 31.4 & 37.6 & 32.3 & 33.8 & 74.4 & 62.8 & 25.0 & \multicolumn{1}{c|}{40.8} & 40.7 \\
        & \textbf{GPT-5 (2025-08-07)} & 33.2 & 40.2 & 30.8 & 37.2 & 78.9 & 61.1 & 25.3 & \multicolumn{1}{c|}{45.2} & 42.2 \\
        & \textbf{GPT-5.1 (2025-11-13)} & 31.0 & 42.9 & 31.3 & \cellcolor[HTML]{DAE8FC}{41.1} & 74.5 & 63.3 & 26.6 & \multicolumn{1}{c|}{41.9} & 42.6 \\
        & \textbf{Gemini-2.5-Flash (2025-06-17)} & 31.8 & 43.3 & 30.9 & 34.2 & 73.0 & 64.4 & 26.4 & \multicolumn{1}{c|}{46.1} & 42.5 \\
        \multirow{-7}{*}{\textbf{\begin{tabular}[c]{@{}l@{}}Proprietary\\ Models\end{tabular}}} & \textbf{Gemini-3.0-Pro-Preview (2025-11-18)} & \cellcolor[HTML]{DAE8FC}34.8 & 48.0 & 32.7 & \cellcolor[HTML]{DAE8FC}{46.1} & 82.0 & \cellcolor[HTML]{DAE8FC}73.6 & 29.2 & \multicolumn{1}{c|}{48.1} & 47.7 \\ \bottomrule
        \end{tabular}
        \caption{Accuracy of user simulation on various domains of datasets (in percentage).}
        \label{Tab:Performance of User Simulation}
\end{table*}

\subsection{Dataset Construction and Fine-tuning LLMs}
\textbf{Datasets.} For the final dataset construction, we reuse the scenes constructed from each domain in Section \ref{Subsec:Data_Distillation}.
Through empirical testing of different dataset sizes (presented in Section \ref{Subsec:DetailedAnalysis}), we determine that 10,000 samples provide an optimal balance between data quality and computational efficiency.
For evaluation, we randomly sample 1,280 and 5,120 scenes for MIND (with real exposures) and other domains (with synthetic exposures) respectively. We fix the length of exposure list in evaluation set to 5. All evaluation scenes are constructed with randomly sampled users and are excluded before the dataset construction phase to prevent possible data leakage~\citep{Sun2020Eval}.
The final constructed dataset set composition and statistics are provided in Appendix \ref{Append:BuiltDataset}.

\textbf{Fine-tuning LLMs.} We primarily utilize \textit{Llama-3.2-3B-Instruct}, the same model employed during the dataset construction phase, for fine-tuning. To validate the generalization capability of our constructed dataset, we also incorporate \textit{Qwen2.5-3B-Instruct}. We adopt a two-stage fine-tuning process for these LLMs. Initially, Supervised Fine-tuning (SFT) is performed using the \textit{accepted} responses. Subsequently, we conduct preference alignment through Direct Preference Optimization~\citep{rafailovDirectPreference2024}.\footnote{Unless explicitly stated otherwise, all subsequent tests are run using \textit{Llama-3.2-3B-Instruct} as the base model, and fine-tuned through the two-stage SFT+DPO process.} Given that predicted behaviors can be verified through simple rules, we also experiment with Group Relative Policy Optimization (GRPO)~\citep{shaoDeepSeekMathPushing2024}. Detailed information regarding the runtime environment and training configurations are listed in the Appendix \ref{subsec:finetune_config}.

\subsection{Is \sysname\ will aligned with user preference? (RQ1)}
\label{Subsec:Performance}
We begin by evaluating the accuracy of our fine-tuned LLM User Simulator in replicating real user behavior across multiple domains. For all inference process, we set the sampling parameters temperature to 1.0 and \textit{top-p} to 0.9. To account for stochastic variation, we randomly sample 5 outputs for each scene and report the mean accuracy.

Table~\ref{Tab:Performance of User Simulation} presents a comprehensive comparison of various LLM served as user simulators. These fine-tuned LLM user simulators not only outperform the base LLMs and stronger teacher LLMs, but also exceed the performance of the most advanced proprietary LLMs in most of the cases, thereby demonstrating the efficacy of leveraging user feedback to better align with user preferences. Moreover, DPO delivers more pronounced improvements compared to GRPO, which we attribute to the relatively simple reward definition employed in GRPO, which could be further explored in the future to fully leverage the potential of reinforcement learning.

\subsubsection{Detailed analysis}
\label{Subsec:DetailedAnalysis}
In this section, we present a detailed analysis on the effect of dataset construction settings on the performance of the fine-tuned user simulators.
\begin{figure}[t!]
    \centering
    \includegraphics[width=\linewidth]{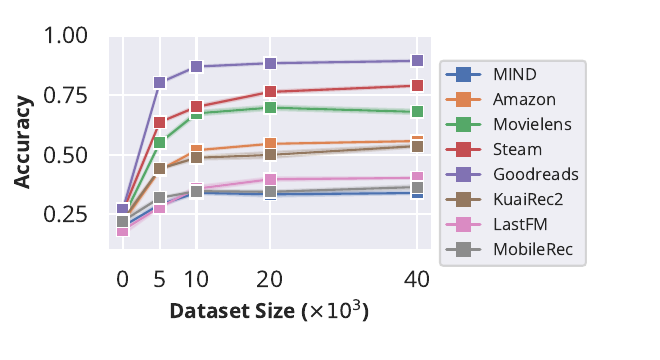}
    \caption{Effect of training dataset size on the performance of user simulator.}
    \label{Fig:Dataset_Size}
\end{figure}

\begin{table}[t]
    \centering
    \renewcommand{\arraystretch}{0.9}
    \scriptsize
    \setlength\tabcolsep{2pt}
    \setul{1pt}{0.4pt}
    \begin{tabular}{lcc}
        \toprule
        \textbf{Dataset} & \textbf{MIND} & \textbf{\begin{tabular}[c]{@{}c@{}}Synthetic Exposure Overall\end{tabular}} \\ \midrule
        \textbf{Llama-3.2-3B-Instruct} & 19.9$_{(0.6)}$ & 22.7$_{(0.5)}$ \\
        \textbf{Random (w/o Decisions)} & 25.7$_{(0.8)}$ & 48.3$_{(0.6)}$ \\
        \textbf{Random (w/ Decisions)} & 31.4$_{(1.4)}$ & 53.9$_{(0.4)}$ \\\midrule
        \textbf{High Accuracy} & 32.4$_{(0.2)}$ & 53.3$_{(0.1)}$ \\
        \textbf{Low Accuracy} & 30.9$_{(0.8)}$ & 54.5$_{(0.2)}$ \\
        \textbf{Diff. Accuracy} & 30.1$_{(0.9)}$ & 53.9$_{(0.2)}$ \\
        \textbf{IFD Score} & 29.9$_{(0.6)}$ & 52.7$_{(0.3)}$ \\ \midrule
        \cellcolor[HTML]{F8CECC}\ \textbf{Ours} & \cellcolor[HTML]{F8CECC}\textbf{34.0$_{(0.6)}$} & \cellcolor[HTML]{F8CECC}\textbf{55.0$_{(0.3)}$} \\ \bottomrule
        \end{tabular}
        \begin{center}
        \begin{tabular}{@{}l@{}}
            *Standard deviations are shown in parentheses. \\ *Accuracy and standard deviation are shown in percentage.
        \end{tabular}
        \end{center}
        \caption{Effect of data selection strategies.}
        \label{Tab:Effect of Data Selection}
\end{table}

\begin{table*}[ht]
    \centering
    \renewcommand{\arraystretch}{1.0}
    \scriptsize
    \setlength\tabcolsep{4pt}
    \setul{1pt}{0.6pt}
    \begin{tabular}{cc|cccccc|cccccc}
        \toprule
        \multicolumn{2}{c|}{} & \multicolumn{6}{c|}{\textbf{Movielens}} & \multicolumn{6}{c}{\textbf{Steam}} \\ \cmidrule(l){3-14} 
        \multicolumn{2}{c|}{\multirow{-2}{*}{\textbf{Model}}} & \textbf{R@5} & \textbf{R@10} & \textbf{N@5} & \textbf{N@10} & \textbf{M@5} & \textbf{M@10} & \textbf{R@5} & \textbf{R@10} & \textbf{N@5} & \textbf{N@10} & \textbf{M@5} & \textbf{M@10} \\ \midrule
         & \textbf{Backbone} & \cellcolor[HTML]{D5E8D4}5.178 & \cellcolor[HTML]{D5E8D4}8.782 & \cellcolor[HTML]{D5E8D4}3.230 & \cellcolor[HTML]{D5E8D4}4.386 & \cellcolor[HTML]{D5E8D4}2.596 & \cellcolor[HTML]{D5E8D4}3.069 & \cellcolor[HTML]{D5E8D4}3.371 & \cellcolor[HTML]{D5E8D4}5.447 & \cellcolor[HTML]{D5E8D4}2.090 & \cellcolor[HTML]{D5E8D4}2.762 & \cellcolor[HTML]{D5E8D4}1.672 & \cellcolor[HTML]{D5E8D4}1.950 \\
         & \textbf{Base LLM} & \cellcolor[HTML]{FFF2CC}5.596 & \cellcolor[HTML]{FFF2CC}9.465 & \cellcolor[HTML]{FFF2CC}3.513 & \cellcolor[HTML]{FFF2CC}4.751 & \cellcolor[HTML]{FFF2CC}2.834 & \cellcolor[HTML]{FFF2CC}3.337 & \cellcolor[HTML]{FFF2CC}3.524 & \cellcolor[HTML]{FFF2CC}5.800 & \cellcolor[HTML]{FFF2CC}2.225 & \cellcolor[HTML]{FFF2CC}3.030 & \cellcolor[HTML]{FFF2CC}1.803 & \cellcolor[HTML]{FFF2CC}2.132 \\
         & \textbf{SFT+DPO} & \cellcolor[HTML]{F8CECC}\textbf{6.089} & \cellcolor[HTML]{F8CECC}\textbf{9.996} & \cellcolor[HTML]{F8CECC}\textbf{3.845} & \cellcolor[HTML]{F8CECC}\textbf{5.098} & \cellcolor[HTML]{F8CECC}\textbf{3.112} & \cellcolor[HTML]{F8CECC}\textbf{3.623} & \cellcolor[HTML]{F8CECC}\textbf{3.707} & \cellcolor[HTML]{F8CECC}\textbf{5.922} & \cellcolor[HTML]{F8CECC}\textbf{2.389} & \cellcolor[HTML]{F8CECC}\textbf{3.108} & \cellcolor[HTML]{F8CECC}\textbf{1.958} & \cellcolor[HTML]{F8CECC}\textbf{2.256} \\ [-1pt]\cmidrule(l){2-14} 
        \multirow{-4}{*}{\textbf{LightGCN}} & \textbf{Improve} & 17.6\% & 13.8\% & 19.0\% & 16.2\% & 19.8\% & 18.0\% & 10.0\% & 8.7\% & 14.3\% & 12.5\% & 17.1\% & 15.7\% \\ [-1pt]\midrule
         & \textbf{Backbone} & \cellcolor[HTML]{D5E8D4}5.197 & \cellcolor[HTML]{D5E8D4}9.219 & \cellcolor[HTML]{D5E8D4}3.313 & \cellcolor[HTML]{D5E8D4}4.613 & \cellcolor[HTML]{D5E8D4}2.700 & \cellcolor[HTML]{D5E8D4}3.236 & \cellcolor[HTML]{D5E8D4}3.195 & \cellcolor[HTML]{D5E8D4}5.337 & \cellcolor[HTML]{D5E8D4}2.010 & \cellcolor[HTML]{D5E8D4}2.693 & \cellcolor[HTML]{D5E8D4}1.622 & \cellcolor[HTML]{D5E8D4}1.899 \\
         & \textbf{Base LLM} & \cellcolor[HTML]{FFF2CC}6.051 & \cellcolor[HTML]{FFF2CC}9.674 & \cellcolor[HTML]{FFF2CC}4.026 & \cellcolor[HTML]{FFF2CC}5.198 & \cellcolor[HTML]{FFF2CC}3.366 & \cellcolor[HTML]{FFF2CC}3.851 & \cellcolor[HTML]{FFF2CC}3.619 & \cellcolor[HTML]{F8CECC}\textbf{5.915} & \cellcolor[HTML]{FFF2CC}2.470 & \cellcolor[HTML]{FFF2CC}3.209 & \cellcolor[HTML]{FFF2CC}2.095 & \cellcolor[HTML]{FFF2CC}2.398 \\
         & \textbf{SFT+DPO} & \cellcolor[HTML]{F8CECC}\textbf{6.449} & \cellcolor[HTML]{F8CECC}\textbf{10.338} & \cellcolor[HTML]{F8CECC}\textbf{4.544} & \cellcolor[HTML]{F8CECC}\textbf{5.793} & \cellcolor[HTML]{F8CECC}\textbf{3.924} & \cellcolor[HTML]{F8CECC}\textbf{4.434} & \cellcolor[HTML]{F8CECC}\textbf{3.619} & \cellcolor[HTML]{FFF2CC}5.886 & \cellcolor[HTML]{F8CECC}\textbf{2.515} & \cellcolor[HTML]{F8CECC}\textbf{3.243} & \cellcolor[HTML]{F8CECC}\textbf{2.155} & \cellcolor[HTML]{F8CECC}\textbf{2.452} \\ [-1pt]\cmidrule(l){2-14} 
        \multirow{-4}{*}{\textbf{DiffRec}} & \textbf{Improve} & 24.1\% & 12.1\% & 37.1\% & 25.6\% & 45.3\% & 37.0\% & 13.3\% & 10.3\% & 25.2\% & 20.4\% & 32.9\% & 29.1\% \\[-1pt]\midrule
         & \textbf{Backbone} & \cellcolor[HTML]{D5E8D4}7.303 & \cellcolor[HTML]{D5E8D4}12.785 & \cellcolor[HTML]{D5E8D4}4.279 & \cellcolor[HTML]{D5E8D4}6.038 & \cellcolor[HTML]{D5E8D4}3.297 & \cellcolor[HTML]{D5E8D4}4.015 & \cellcolor[HTML]{FFF2CC}5.191 & \cellcolor[HTML]{FFF2CC}8.949 & \cellcolor[HTML]{FFF2CC}3.268 & \cellcolor[HTML]{FFF2CC}4.475 & \cellcolor[HTML]{FFF2CC}2.640 & \cellcolor[HTML]{FFF2CC}3.134 \\
         & \textbf{Base LLM} & \cellcolor[HTML]{FFF2CC}8.024 & \cellcolor[HTML]{FFF2CC}13.961 & \cellcolor[HTML]{FFF2CC}4.800 & \cellcolor[HTML]{FFF2CC}6.689 & \cellcolor[HTML]{FFF2CC}3.748 & \cellcolor[HTML]{FFF2CC}4.510 & \cellcolor[HTML]{D5E8D4}4.621 & \cellcolor[HTML]{D5E8D4}8.927 & \cellcolor[HTML]{D5E8D4}2.836 & \cellcolor[HTML]{D5E8D4}4.212 & \cellcolor[HTML]{D5E8D4}2.256 & \cellcolor[HTML]{D5E8D4}2.814 \\
         & \textbf{SFT+DPO} & \cellcolor[HTML]{F8CECC}\textbf{8.365} & \cellcolor[HTML]{F8CECC}\textbf{13.999} & \cellcolor[HTML]{F8CECC}\textbf{5.027} & \cellcolor[HTML]{F8CECC}\textbf{6.852} & \cellcolor[HTML]{F8CECC}\textbf{3.938} & \cellcolor[HTML]{F8CECC}\textbf{4.694} & \cellcolor[HTML]{F8CECC}\textbf{5.637} & \cellcolor[HTML]{F8CECC}\textbf{9.549} & \cellcolor[HTML]{F8CECC}\textbf{3.606} & \cellcolor[HTML]{F8CECC}\textbf{4.864} & \cellcolor[HTML]{F8CECC}\textbf{2.945} & \cellcolor[HTML]{F8CECC}\textbf{3.460} \\ [-1pt]\cmidrule(l){2-14} 
        \multirow{-4}{*}{\textbf{SASRec}} & \textbf{Improve} & 14.5\% & 9.5\% & 17.5\% & 13.5\% & 19.4\% & 16.9\% & 8.6\% & 6.7\% & 10.3\% & 8.7\% & 11.6\% & 10.4\% \\ [-1pt]\midrule
         & \textbf{Backbone} & \cellcolor[HTML]{D5E8D4}6.791 & \cellcolor[HTML]{D5E8D4}12.747 & \cellcolor[HTML]{D5E8D4}3.973 & \cellcolor[HTML]{D5E8D4}5.879 & \cellcolor[HTML]{D5E8D4}3.059 & \cellcolor[HTML]{D5E8D4}3.834 & \cellcolor[HTML]{D5E8D4}5.272 & \cellcolor[HTML]{D5E8D4}9.074 & \cellcolor[HTML]{D5E8D4}3.352 & \cellcolor[HTML]{D5E8D4}4.573 & \cellcolor[HTML]{FFF2CC}2.725 & \cellcolor[HTML]{D5E8D4}3.223 \\
         & \textbf{Base LLM} & \cellcolor[HTML]{FFF2CC}8.024 & \cellcolor[HTML]{FFF2CC}14.662 & \cellcolor[HTML]{FFF2CC}4.761 & \cellcolor[HTML]{FFF2CC}6.891 & \cellcolor[HTML]{F8CECC}\textbf{3.700} & \cellcolor[HTML]{F8CECC}\textbf{4.569} & \cellcolor[HTML]{FFF2CC}5.454 & \cellcolor[HTML]{FFF2CC}9.681 & \cellcolor[HTML]{FFF2CC}3.389 & \cellcolor[HTML]{FFF2CC}4.741 & \cellcolor[HTML]{D5E8D4}2.716 & \cellcolor[HTML]{FFF2CC}3.265 \\
         & \textbf{SFT+DPO} & \cellcolor[HTML]{F8CECC}\textbf{8.175} & \cellcolor[HTML]{F8CECC}\textbf{14.757} & \cellcolor[HTML]{F8CECC}\textbf{4.772} & \cellcolor[HTML]{F8CECC}\textbf{6.895} & \cellcolor[HTML]{FFF2CC}3.672 & \cellcolor[HTML]{FFF2CC}4.545 & \cellcolor[HTML]{F8CECC}\textbf{5.783} & \cellcolor[HTML]{F8CECC}\textbf{9.907} & \cellcolor[HTML]{F8CECC}\textbf{3.716} & \cellcolor[HTML]{F8CECC}\textbf{5.037} & \cellcolor[HTML]{F8CECC}\textbf{3.037} & \cellcolor[HTML]{F8CECC}\textbf{3.575} \\ [-1pt]\cmidrule(l){2-14} 
        \multirow{-4}{*}{\textbf{NARM}} & \textbf{Improve} & 20.4\% & 15.8\% & 20.1\% & 17.3\% & 20.0\% & 18.5\% & 9.7\% & 9.2\% & 10.9\% & 10.1\% & 11.4\% & 10.9\% \\ [-1pt] \bottomrule
        \end{tabular}
    \caption{Performance comparison of user simulators on recommendation scenarios. (In percentage)}
    \label{Tab:Rec_Improvement}
    \end{table*}

\textbf{Effect of training dataset size.} To validate the size of datasets on the performance of fine-tuned LLM simulator, we construct a series of variants using our pipeline, with the varying number from 5,000 to 40,000. We increase the epoches when smaller datasets is used to mitigate the impact of inadequate training. As shown in Figure \ref{Fig:Dataset_Size}, accuracy rises sharply until approximately 10,000 samples, after which further increases yield only marginal improvements. It indicates that, beyond a certain dataset size, additional examples contribute limited gains in preference alignment and generalization, as existing data are already able to provide sufficient preference alignment information.

\textbf{Effects of data selection strategies.} To validate the effect of data selection strategy in constructing high-quality dataset, we compare a range of baselines: random sampling with or without decision processes, selection based on high, low, or differential prediction accuracy of the predicted behaviors, and selecting using the Instruction-Following Difficulty (IFD)~\citep{liSuperfilteringWeaktoStrong2024} score. (Detailed setup of baselines is listed in Appendix \ref{Append:Ablation_Setup}.)
In our pipeline, as summarized in Table \ref{Tab:Effect of Data Selection}, our dataset construction pipeline outperforms these baselines across both real and synthetic exposures. This result confirms our pipeline, which combines uncertainty-based measures and behavior sampling based on real user feedback, yields robust, preference-aligned simulators than accuracy-based or instruction-following filtering metrics.

\subsubsection{Analysis of decision making process}    
We conduct an in-depth analysis of how the decision-making process generated by our user simulator affects the accuracy of user behavior prediction. We require the model to summarize the key factors in each stage of decision-making process. We extract these factors from the generated results on the test set, then clean and categorize them, merging semantically similar terms. Ultimately, we derive factors from 3 stages: \textit{Stimulus}, \textit{Knowledge}, and \textit{Evaluation}.

\begin{figure}[t!]
    \centering
    \begin{subfigure}[b]{0.49\linewidth}
        \centering
        \includegraphics[width=\linewidth]{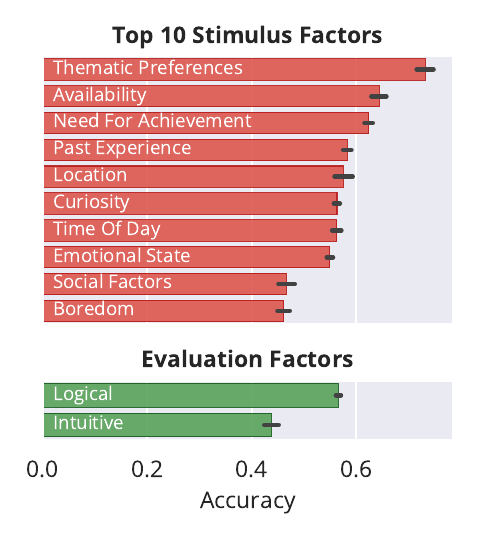}
        \caption{Content}
        \label{Fig:Acc_Factors}
    \end{subfigure}
    \hfill
    \begin{subfigure}[b]{0.49\linewidth}
        \centering
        \includegraphics[width=\linewidth]{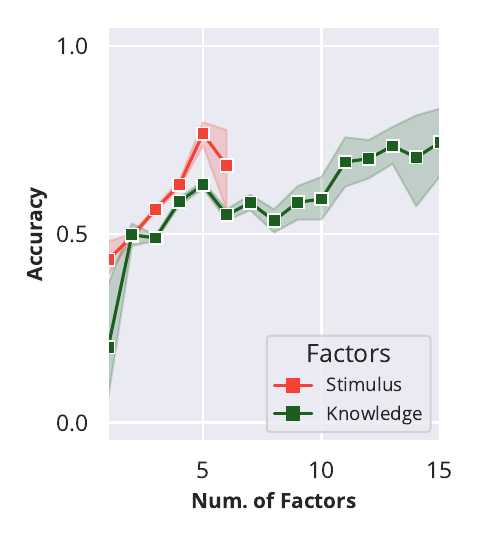}
        \caption{Number}
        \label{Fig:Len_Factors}
    \end{subfigure}
    \caption{Effect of the generated decision-making process on the performance of user simulator.}
\end{figure}

Our statistical analysis reveals that both the stimulus factors and the final evaluation style are correlated with prediction accuracy. As shown in Figure \ref{Fig:Acc_Factors}, when users simulator explicitly express \textit{"thematic preferences"}, the model achieves significantly higher prediction accuracy compared to cases where only \textit{"boredom"} are expressed. Additionally, decision processes adopt a \textit{"logical"} style tend to rely more on the LLM's reasoning ability, yielding higher accuracy than \textit{"intuitive"} style.
"
We also find a positive correlation between the length of descriptions for stimulus and knowledge factors and the model's prediction accuracy, as shown in Figure \ref{Fig:Len_Factors}. In other words, the more factors a user considers and the more detailed the descriptions during the decision process, the better the alignment between the model's predictions and actual user behavior. This finding suggests that rich and diverse decision process information helps improve the performance of the user simulator.

In summary, the generative decision-making process provides greater interpretability for user modeling and valuable intermediate variables for subsequent behavior analysis and causal inference.

\subsection{Is \sysname~ helpful in providing feedback to RS? (RQ2)}
To evaluate whether our user simulator can provide valuable feedback signals to RS, we design an interactive simulation framework to run the test. First, we train a RS on the original training dataset (labeled as \textbf{"Backbone"}), where all user interactions used in fine-tuning the simulator are excluded. Next, we generate a candidate recommendation list and expose it to the user simulator, collecting its simulated feedback. These synthetic interactions are then appended to the RS's training data, and the model is retrained to incorporate the additional signals. We conduct this procedure across 10 representative RS algorithms (Due to space limitations, we present the results of LightGCN~\citep{heLightGCNSimplifying2020}, DiffRec~\citep{wangDiffusionRecommender2023}, SASRec~\citep{kangSelfAttentiveSequential2018}, and NARM~\citep{Li2017NeuralAtt}, in Table~\ref{Tab:Rec_Improvement}. Additional model results are provided in Appendix ~\ref{Appen:RSInteraction}.) based on two datasets Movielens and Steam, comparing the performance of the fine-tuned simulator against the base LLM. We evaluate each scenario using \textit{Recall (R@k)}, \textit{Normalized Discounted Cumulative Gain (N@k)} and \textit{Mean Reciprocal Rank (M@k}), where $k$ denotes denotes the number of items recommended to the user by the RS.

As shown in Table \ref{Tab:Rec_Improvement}, augmenting RS training with feedback from our fine-tuned simulator yields consistent gains in all three metrics, demonstrating that preference-aligned simulation produces constructive signals that significantly enhance recommendation quality.


\section{Conclusion and Future Work}
In this work, we address the challenges of ambiguity, noise, and scale in leveraging user feedback for user simulation in RSs. Our proposed framework, \sysname, fully exploits user feedback to align LLM-based user simulators with authentic user preferences. It transforms raw feedback into unified simulation scenarios, generates explicit human-like decision-making processes, and distills high-quality examples for fine-tuning. Extensive experiments demonstrate that our fine-tuned simulators achieve superior alignment with real user behavior and provide more interpretable feedback to RSs, resulting in significant improvements in recommendation quality.

More broadly, the cognitive modeling perspective underlying our work aligns with the digital twins paradigm in AI-driven decision systems~\citep{aliCognitiveDigitalTwins2021}. 
Our approach builds virtual user replicas by extracting explicit reasoning from noisy behavior, similar to constructing digital counterparts in other domains using imperfect real-world signals. For example, in battery health management, physics-informed models combine domain knowledge and data-driven learning to predict degradation from noisy sensor data under uncertainty~\citep{zhuPhysicsInformedMultiTask2026}.
Both settings use knowledge transfer to bridge rich expert models and lightweight deployable replicas, and leverage uncertainty-aware selection to address observation noise~\citep{kangStatisticalUncertainty2021,liUncertaintyAwareDecision2025}.
These parallels suggest our methodology may also benefit other AI systems requiring reliable decision-making under noisy, incomplete observations.

\section*{Limitations}
Our work has several limitations that point to promising directions for future research. First, our current framework does not incorporate multi-modal information (e.g., images, audio, or video), which can be crucial for user simulation in domains such as short video or music recommendation. Second, our approach focuses on modeling explicit user interactions and does not address the scenario where users are exposed to items but choose not to interact (i.e., user leaving or non-interaction feedback). While some prior work introduce modules to predict such user-leaving behaviors, the lack of corresponding public data makes it challenging to construct and evaluate these cases in our setting. As a result, our study centers on leveraging existing explicit feedback to enhance user simulation, and we leave the implicit non-interaction modeling for future work.

\section*{Ethical Considerations}
User simulators trained on recommendation feedback may inherit the biases embedded in the underlying data. For instance, skewed preference distributions across demographic groups or content domains can be amplified by the simulator. To mitigate this risk, we deliberately incorporated feedback from as many diverse domains as possible into the training process, aiming to reduce systematic bias and improve generalizability.

Although the simulator leverages user-level profiles, we exclusively rely on publicly available datasets in which the original authors have already applied rigorous anonymization procedures. Consequently, no personally identifiable information is retained. Any resemblance of artificial generated profile descriptions and decision making process to real individuals is purely coincidental. All dataset reuse, modification, and related operations strictly complied with the copyright statements and usage policies issued by the respective data owners.

\section*{Acknowledgments}
This research is supported by A*STAR under its MTC Programmatic (Award M23L9b0052); and by the Ministry of Education, Singapore, under its Academic Research Fund (AcRF) Tier 1 grant, and funded through the SMU-SUTD Internal Research Grant Call (SMU-SUTD 2023\_02\_01); and by the Ministry of Education, Singapore, under its Academic Research Fund Tier 2 (Award No. MOE-T2EP201230015).
\bibliography{Dataset,UserMirrorer}

\newpage
\appendix
\newpage

\section{Dataset}

\subsection{Statistics of Source User Feedback Datasets in RS}
\label{subsec:appendix_dataset}

\begin{table*}[h]
\renewcommand{\arraystretch}{0.8}\
\setlength\tabcolsep{2pt}
\centering
\scriptsize
  \begin{tabular}{@{}cl|cccccccccc@{}}
\toprule
\multicolumn{2}{c|}{\multirow{2}{*}{\textbf{Dataset}}} & \multirow{2}{*}{\textbf{Domain}} & \multirow{2}{*}{\textbf{\begin{tabular}[c]{@{}c@{}}User\\ Feature\end{tabular}}} & \multirow{2}{*}{\textbf{\begin{tabular}[c]{@{}c@{}}Item\\ Feature\end{tabular}}} & \multirow{2}{*}{\textbf{\begin{tabular}[c]{@{}c@{}}Behavior\\ Feature\end{tabular}}} & \multicolumn{3}{c}{\textbf{Additional Information}} & \multicolumn{3}{c}{\textbf{Dataset Details}} \\ \cmidrule(l){7-12} 
\multicolumn{2}{c|}{} &  &  &  &  & \textbf{Exposure} & \textbf{Review} & \textbf{Social} & \textbf{\#User} & \textbf{\#Item} & \textbf{\#Interaction} \\ \midrule
\multicolumn{2}{c|}{\textbf{MIND}} & News &  & \checkmark & \checkmark & \checkmark &  &  & 307,353 & 130,379 & 552,043 \\
\multicolumn{2}{c|}{\textbf{MovieLens}} & Movie & \checkmark & \checkmark & \checkmark &  &  &  & 6,040 & 3,883 & 1,000,209 \\
\multicolumn{2}{c|}{\textbf{KuaiRec2}} & Short video & \checkmark & \checkmark & \checkmark &  &  & \checkmark & 7,176 & 10,728 & 12,530,806 \\
\multicolumn{2}{c|}{\textbf{Goodreads}} & Book &  & \checkmark & \checkmark &  & \checkmark &  & 342,415 & 89,411 & 6,995,891 \\
\multicolumn{2}{c|}{\textbf{MobileRec}} & APPs &  & \checkmark & \checkmark &  &  &  & 15,141 & 10,173 & 100,000 \\
\multicolumn{2}{c|}{\textbf{Steam}} & Game & \checkmark & \checkmark & \checkmark &  & \checkmark &  & 250,793 & 32,132 & 2,091,415 \\
\multicolumn{2}{c|}{\textbf{LastFM}} & Music &  & \checkmark & \checkmark &  &  & \checkmark & 1,892 & 17,632 & 70,297 \\ \midrule
\multirow{4}{*}{\textbf{Amazon}} & \textbf{Fashion} & E-commerce &  & \checkmark & \checkmark &  & \checkmark &  & 14,791 & 97,287 & 56,086 \\
 & \textbf{Beauty} & E-commerce &  & \checkmark & \checkmark &  & \checkmark &  & 1,620 & 37,580 & 14,984 \\
 & \textbf{Grocery} & E-commerce &  & \checkmark & \checkmark &  & \checkmark &  & 24,076 & 63,122 & 378,775 \\
 & \textbf{Office} & E-commerce &  & \checkmark & \checkmark &  & \checkmark &  & 11,453 & 64,726 & 162,297 \\ \bottomrule
\end{tabular}
\caption{Descriptions of Datasets.}
\label{tab:details_of_trained_model}
\end{table*}

In this work, we collect multiple open-source user feedback datasets from various domains to construct \sysname. 
Our dataset selection is based on two key criteria: (1) coverage across multiple recommendation domains to enhance the generalization ability of LLMs across different recommendation scenarios, and (2) the inclusion of human-readable item descriptions to align with human cognition.
The detailed descriptions of the selected datasets are as follows:

\begin{itemize}[leftmargin=12pt] 
    \item \textbf{Microsoft News Dataset (MIND)}~\citep{wuMINDLargescale2020} is a large-scale news recommendation dataset from \textit{Microsoft News}\footnote{\url{https://microsoftnews.msn.com/}}, containing user interactions (impressions, clicks) and metadata (titles, abstracts and categories). 
    It is widely used for studying personalized news recommendations.
    \item \textbf{MovieLens-1M (MovieLens)}~\citep{harper2015movielens} is a popular dataset for collaborative filtering, featuring around 1 million explicit movie ratings (1–5 scale) from the \textit{MovieLens}\footnote{\url{https://movielens.org/}} platform. 
    It is commonly used for evaluating rating prediction algorithms and user preference modeling. 
    \item \textbf{KuaiRec2}~\citep{gao2022kuairec} is a large-scale dataset from \textit{Kuaishou}\footnote{\url{https://www.kuaishou.com/}}, a short-video platform, containing user interactions such as watch behavior, likes, and shares. This data follows CC-BY-SA 4.0 license.
    It is valuable for studying sequential recommendations and user engagement in short-video scenarios.
    \item \textbf{Goodreads}~\citep{wan2018item} is a book recommendation dataset from \textit{Goodreads}\footnote{\url{https://www.goodreads.com/}}, featuring user ratings, reviews, and reading behaviors. 
    It is useful for analyzing long-term preferences, content-based recommendations, and social influence. The data follows the Apache 2.0 License.
    \item \textbf{MobileRec}~\citep{maqbool2023mobilerec} is a mobile app recommendation dataset that records user interactions, including installs, usage duration, and engagement. 
    It supports research on context-aware recommendations and user experience optimization in mobile environments.
    \item \textbf{Steam}~\citep{o2016condensing} is a game recommendation dataset from the \textit{Steam}\footnote{\url{https://store.steampowered.com/}} platform, containing user purchase history, playtime, and reviews. It is useful for studying implicit feedback and analyzing the impact of game popularity on user engagement.  
    \item \textbf{LastFM}~\citep{Bertin-Mahieux2011} is a music recommendation dataset from \textit{Last.fm}\footnote{\url{https://www.last.fm/}}, capturing user listening history, artist preferences, and tagging behaviors. It supports research on collaborative filtering, content-based recommendations, and dynamic user preference modeling.
    \item \textbf{Amazon}~\citep{ni2019justifying} is a large-scale e-commerce dataset covering user-item interactions across multiple domains from the e-commerce platform \textit{Amazon\footnote{\url{https://www.amazon.com/}}}. It is widely used for evaluating product recommendation models. Key domain-specific subsets include:  
    \begin{itemize}[leftmargin=12pt]  
        \item \textbf{Fashion}: User purchase histories and reviews for clothing, accessories, and footwear.  
        \item \textbf{Beauty}: Interactions with beauty and personal care products.  
        \item \textbf{Grocery}: User transactions and ratings in the grocery sector.  
        \item \textbf{Office}: Interactions related to office supplies, electronics, and similar items.  
    \end{itemize}  
    These subsets support research on cross-domain recommendations, personalized ranking, and user behavior modeling in e-commerce. 
\end{itemize}

\subsection{Details of the Construction of High-quality User Simulation Dataset}
\label{Append:BuiltDataset}
We introduced a framework \sysname~for data distillation in which we sample user simulation scenes and generate decision-making processes to construct a high-quality dataset. For our experiments, we randomly sampled 16,384 user scenes for each domain (except for Amazon-Beauty which has less than 16,384). We then employed two large language models with contrasting capacities: a stronger LLM \textit{Qwen-2.5-32B-Instruct} and a lightweight LLM \textit{Llama-3.2-3B-Instruct}. All inferences were conducted via the \textit{vLLM} framework, producing $N = 10$ candidate outputs with the sampling parameter temperature set to 1.0 and top-p set to 0.9. After filtering out any samples that exhibited generation errors, we record the valid sample counts and total token consumption, as summarize in the Table~\ref{Tab:tokens} below. The resulting distilled dataset comprises 10,000 simulated decision-making scenes for training and 6,400 scenes for testing. Table \ref{tab:construct_dataset} lists the statistics of the training set and test set.

All data is completely artificially generated. Any similarity in profile descriptions to actual persons, living or dead, is purely coincidental. Our data follows CC-BY-SA 4.0 license. All reuse, modification, and related operations on these datasets strictly adhered to the copyright statements of the data owners. 

\begin{table*}[ht]
    \scriptsize
    \renewcommand{\arraystretch}{1.0}\
    \setlength\tabcolsep{4pt}
    \centering
    \setul{1pt}{0.4pt}
    \begin{tabular}{@{}l|c|cccc|cccc@{}}
        \toprule
        \multirow{3}{*}{\textbf{Dataset}} & \multirow{3}{*}{\textbf{\# Scene}} & \multicolumn{4}{c|}{\textbf{Qwen-2.5-32B-Instruct}} & \multicolumn{4}{c}{\textbf{Llama-3.2-3B-Instruct}} \\ \cmidrule(l){3-10} 
         &  & \multicolumn{2}{c}{\textbf{Input}} & \multicolumn{2}{c|}{\textbf{Output}} & \multicolumn{2}{c}{\textbf{Input}} & \multicolumn{2}{c}{\textbf{Output}} \\ \cmidrule(l){3-10} 
         &  & \textbf{Average} & \multicolumn{1}{l}{\textbf{Total (M)}} & \textbf{Average} & \multicolumn{1}{l|}{\textbf{Total (M)}} & \textbf{Average} & \multicolumn{1}{l}{\textbf{Total (M)}} & \textbf{Average} & \multicolumn{1}{l}{\textbf{Total (M)}} \\ \midrule
        \textbf{KuaiRec2} & 16,384 & 1,847 & 30.3 & 237 & 38.9 & 1,902 & 31.2 & 245 & 40.1 \\
        \textbf{LastFM} & 16,384 & 1,111 & 18.2 & 231 & 37.9 & 1,126 & 18.5 & 243 & 39.8 \\
        \textbf{MobileRec} & 16,384 & 1,455 & 23.8 & 244 & 40.0 & 1,435 & 23.5 & 267 & 43.7 \\
        \textbf{Amazon-Beauty} & 10,124 & 1,344 & 13.6 & 247 & 25.0 & 1,345 & 13.6 & 273 & 27.7 \\
        \textbf{Amazon-Fashion} & 16,384 & 1,247 & 20.4 & 235 & 38.5 & 1,252 & 20.5 & 262 & 42.9 \\
        \textbf{Amazon-Office} & 16,384 & 1,623 & 26.6 & 254 & 41.6 & 1,595 & 26.1 & 273 & 44.7 \\
        \textbf{Amazon-Grocery} & 16,384 & 1,572 & 25.8 & 238 & 39.0 & 1,553 & 25.5 & 254 & 41.6 \\
        \textbf{Movielens} & 16,384 & 1,538 & 25.2 & 252 & 41.2 & 1,498 & 24.5 & 266 & 43.5 \\
        \textbf{Steam} & 16,383 & 2,315 & 37.9 & 262 & 43.0 & 2,268 & 37.2 & 290 & 47.5 \\
        \textbf{Goodreads} & 16,384 & 2,694 & 44.1 & 267 & 43.7 & 2,609 & 42.7 & 290 & 47.5 \\
        \textbf{MIND} & 16,335 & 1,129 & 18.4 & 178 & 29.1 & 1,142 & 18.7 & 279 & 45.6 \\ \midrule
        \textbf{Overall} & \textbf{173,914} & \textbf{1,625} & \textbf{284.4} & \textbf{241} & \textbf{417.9} & \textbf{1,611} & \textbf{281.9} & \textbf{267} & \textbf{464.7} \\ \bottomrule
        \end{tabular}
        \caption{Token usage during the dataset construction phase.}
        \label{Tab:tokens}
    \end{table*}

\begin{table*}[ht]
    \footnotesize
    \renewcommand{\arraystretch}{1.0}\
    \setlength\tabcolsep{3pt}
    \centering
    \scriptsize
    \setul{1pt}{0.4pt}
    \begin{tabular}{@{}lcccccccccccc@{}}
        \toprule
        \multirow{2}{*}{\textbf{Dataset}} & \multirow{2}{*}{\textbf{Goodreads}} & \multirow{2}{*}{\textbf{KuaiRec2}} & \multirow{2}{*}{\textbf{Steam}} & \multirow{2}{*}{\textbf{Movielens}} & \multirow{2}{*}{\textbf{LastFM}} & \multirow{2}{*}{\textbf{MobileRec}} & \multicolumn{4}{c}{\textbf{Amazon}} & \multirow{2}{*}{\textbf{MIND}} & \multirow{2}{*}{\textbf{Overall}} \\ \cmidrule(lr){8-11}
         &  &  &  &  &  &  & Grocery & Beauty & Fashion & Office &  &  \\ \midrule
        \textbf{\# Training} & 1,184 & 1,093 & 1,055 & 1,022 & 737 & 657 & 985 & 570 & 910 & 823 & 964 & \textbf{10,000} \\ \midrule
        \textbf{\# Test} & 640 & 640 & 640 & 640 & 640 & 640 & 320 & 320 & 320 & 320 & 1,280 & \textbf{6,400} \\ \bottomrule
        \end{tabular}
    \caption{Statistics of constructed datasets.}
    \label{tab:construct_dataset}
    \end{table*}

\section{Details of Simulation Scene and Behavior Construction}
\subsection{Construction of Exposure Lists}
\label{subsec:appendix_exposure_list}
To simulate the decision-making process in which users are exposed to multiple options before making a final selection, we reconstruct an exposure for each recorded user interaction.  
The reconstruction of exposures is achieved by using a combination of context-aware, collaborative, and random strategies, thereby providing LLMs with a more human-aligned decision context.

\textbf{Context-aware Sampling Strategy.} 
This strategy constructs exposure lists by retrieving items that are semantically similar to those the user has previously interacted with. 
Specifically, we utilize \textit{GTE-Qwen2-1.5B-Instruct} model~\citep{liGeneralText2023} to encode the textual content of all items into dense embeddings. 
Then, for a given target item, we compute cosine similarities in the embedding space and retrieve the top-$K$ most similar items to simulate items likely to co-occur in the user's exposure context.

\textbf{Collaborative Sampling Strategy.}
To reflect real-world recommendation scenarios, we employ collaborative filtering signals to model system-driven exposure. 
We first pre-train a recommendation model, \textit{GF-CF}~\citep{shen2021powerful}, to learn latent interaction patterns from the training data. 
Given a user, we use this model to rank all items and select the top-$K$ items as candidates that the system would most likely expose.

\textbf{Random Sampling Strategy.}
To ensure diversity and account for noise or exploration in RSs, we also include a strategy that randomly samples items the user has not previously interacted with. 
This allows the exposure list to include both relevant and irrelevant items, better approximating the uncertain and imperfect nature of real-world exposures.

Once the three candidate lists are generated using the above strategies, we combine them into a unified exposures via a randomized iterative procedure (as shown in Algorithm~\ref{alg:exposure-construction}). 
At each iteration, we randomly select one of the three candidate lists with equal probability and extract its top-ranked item. 
If the item is not yet present in the final list, it is appended. 
This process repeats until the maximum exposure length $N$ is reached. 
To ensure that the ground-truth item (i.e., the item the user actually interacted with) appears in the list, we insert it at a randomly chosen position.
This strategy ensures that the exposure list reflects a mixture of relevance (from context and collaborative cues), diversity (via random sampling), and decision-critical signal (via inclusion of the ground-truth item), thereby supporting more cognitively plausible user simulations.
\begin{algorithm}[ht]
\caption{Exposure List Construction}
\label{alg:exposure-construction}
\KwIn{User $u$, item set $\mathcal{L}$, ground-truth item $l^*$, maximum exposure length $N$}
\KwOut{Final exposure list $\rmL$}

Generate three candidate lists: \\
\Indp
$\rmL_{\text{content}} \leftarrow$ content-based top-$K$ ranking of $\mathcal{L}$ for $u$ \\
$\rmL_{\text{collab}} \leftarrow$ collaborative filtering top-$K$ ranking of $\mathcal{L}$ for $u$ \\
$\rmL_{\text{rand}} \leftarrow$ random $K$ permutation of $\mathcal{L}$ \\
\Indm

Initialize exposure list: $\rmL \leftarrow \emptyset$ \\
Candidate pool: $\mathcal{C} \leftarrow \{\rmL_{\text{content}}, \rmL_{\text{collab}}, \rmL_{\text{rand}}\}$ \\

\While{$|\rmL| < N$}{
    Randomly select a list $\rmL_i \in \mathcal{C}$ with uniform probability \\
    Pop the top-ranked item $l$ from $\rmL_i$ \\
    \If{$l \notin \rmL$}{
        Append $l$ to $\rmL$ \\
    }
}

Randomly select a position $p \in [0, N]$ \\
Insert $l^*$ at position $p$ in $\rmL$ \\

\Return{$\rmL$}
\end{algorithm}

In this work, except for the MIND dataset which contains the real exposures, we construct the synthetic exposure lists for other datasets following the above Algorithm \ref{alg:exposure-construction}.

\subsection{Dataset construction pipeline in \sysname}
\label{Append:DatasetPipeline}
We construct a data selection mechanism to extract high-quality samples from massive raw user feedback data that are more valuable for aligning LLM user simulators with the following steps:
\begin{enumerate}[leftmargin=12pt]
\item \textbf{Simulation Scene Construction}: Randomly sampling from raw user feedback to construct a batch of user simulation scenes.
\item \textbf{Decision Process Generation}: Using LLM $A$ and $B$ with different capabilities to generate $N$ decision processes with predicted behaviors for each sample.
\item \textbf{Uncertainty-based Scene Distillation}: Calculate $\Delta_{EU}(\bm X, (A, B))$ via Equation~\ref{Eq:DiffUncert}, sort scenes by descending $\Delta_{EU}$ values, and sequentially retain those with the highest values.
\item \textbf{Behavior Reject Sampling}: We discard the scene where all behaviors predicted by stronger LLM are unmatched with the behavior in raw data. For the remaining scene, we mark the decision process according to the matching of behaviors as \textit{accepted} and \textit{rejected}, and formulate the samples with highest confidence as the preference pair.
\end{enumerate}
\section{Experimental Configurations}

\subsection{Experiment Environment}
All experiments were conducted on a unified setup featuring an \textit{AMD EPYC 7713} 64-core processor and four \textit{NVIDIA A100} 40GB GPUs. LLM fine-tuning (SFT and DPO stages) was performed using \textit{torchtune}\footnote{https://github.com/pytorch/torchtune}, and GRPO was implemented with \textit{verl}\footnote{https://github.com/volcengine/verl}. For LLM inference, we employed \textit{vllm}\footnote{https://github.com/vllm-project/vllm}, and RecBole~\citep{zhaoRecBoleUnified2021a} was used for training and evaluating the recommendation system models.

\subsection{LLM Fine-tuning Configurations}
\label{subsec:finetune_config}
\begin{table*}[h]
    \renewcommand{\arraystretch}{1.2}\
    \setlength\tabcolsep{2pt}
    \centering
    \scriptsize
        \begin{tabular}{@{}lcccccccc@{}}
            \toprule
            \textbf{Hyperparameter} & \textbf{Epoches} & \textbf{Learning Rate} & \textbf{\begin{tabular}[c]{@{}c@{}}Batch\\ Size\end{tabular}} & \textbf{Optimizer} & \textbf{\begin{tabular}[c]{@{}c@{}}Weight\\ Decay\end{tabular}} & \textbf{\begin{tabular}[c]{@{}c@{}}Decay\\ Schedule\end{tabular}} & \textbf{\begin{tabular}[c]{@{}c@{}}Warm-up\\ Steps\end{tabular}} & \textbf{\begin{tabular}[c]{@{}c@{}}Other\\ Hyperparameter\end{tabular}} \\ \midrule
            \textbf{SFT} & [1,2,3] & [2e-4, 1e-4, 5e-5, 2e-5] & 16 & AdamW & 0.01 & Cosine Decay & 10 &  \\
            \textbf{DPO} & [1,2,3] & [2e-4, 1e-4, 5e-5, 2e-5] & 16 & AdamW & 0.05 & Cosine Decay & 10 & beta: 0.1 \\
            \textbf{GRPO} & [1,2,3] & 1e-5 & 64 & AdamW & 0.01 & Cosine Decay & 10 & N: 8; KL Coeff: 1e-3 \\ \bottomrule
            \end{tabular}
        \caption{Configurations of LLM fine-tuning stages.}
        \label{tab:llm_finetune}
    \end{table*}

To fine-tune our LLMs on the constructed user feedback dataset, we use the hyperparameters in Table \ref{tab:llm_finetune}. Unspecified settings follow the defaults of the \textit{torchtune} and \textit{verl} frameworks.
We consider two training setups:
\begin{itemize}[leftmargin=24pt]
    \item \textbf{Single-Stage SFT:} For models without the decision-making process, we apply supervised fine-tuning only, treating the task as single-token classification.
    \item \textbf{Two-Stage Fine-Tuning:} For models incorporating decision-making, we perform a standard two-stage pipeline: supervised fine-tuning (SFT) followed by decision-focused fine-tuning (e.g., DPO) using our preference data.
\end{itemize}

We also employ the following two strategies in all training process:
\begin{itemize}[leftmargin=24pt]
    \item \textbf{Grid Search:} We systematic explore the learning rates and number of epoches to identify the optimal settings of fine-tuning.
    \item \textbf{Mixture of General Purpose Data:} During SFT, we mix in examples from the \textit{Tulu-v3-mixture}~\citep{lambert2024tulu3} dataset to prevent catastrophic forgetting, maintaining a 10:1 ratio of task-specific to general data.
\end{itemize}

\subsection{Configurations of Experiments on Interactive Simulation with RS}
\label{Appen:RSInteraction}
We conducted a series of experiments to evaluate whether the simulated user behaviors generated by \sysname~can effectively benefit recommendation models. The experimental procedure consists of three main steps:

\textbf{Step 1: Training Backbone Recommendation Models.}
To prevent potential information leakage, we first removed all user data used to train \sysname~from the dataset (Movielens and Steam).
The remaining data was then split into training ($\mathcal{D}_{train}$), validation ($\mathcal{D}_{valid}$), and test sets ($\mathcal{D}_{test}$) with a ratio of 8:1:1. 
We selected eight recommendation models based on diverse technical paradigms and trained them using standard procedures. 
\begin{itemize}
    \item \textbf{BPR}~\citep{rendleBPRBayesian2009}: Bayesian Personalized Ranking is a pairwise matrix factorization model that learns personalized rankings by optimizing the relative order of observed and unobserved items.

    \item \textbf{LightGCN}~\citep{heLightGCNSimplifying2020}: Light Graph Convolution Network simplifies traditional GCNs by discarding feature transformations and nonlinearities, focusing solely on neighbor aggregation to improve efficiency and accuracy.

    \item \textbf{MultiDAE}~\citep{liang2018variational}: The Multinomial Denoising Autoencoder learns robust user representations by reconstructing interaction vectors from corrupted input, effectively modeling non-linear user preferences.

    \item \textbf{DiffRec}~\citep{wangDiffusionRecommender2023}: Diffusion recommendation model formulates recommendation as a denoising diffusion process, introducing noise to user signals and learning to recover them through a multi-step generative model with structural uncertainty.

    \item \textbf{SGL}~\citep{wuSelfsupervisedGraph2021}: Self-supervised Graph Learning augments graph-based collaborative filtering with contrastive learning, enforcing consistency between perturbed views of the user-item interaction graph.

    \item \textbf{NeuMF}~\citep{he2017neural}: Neural Matrix Factorization combines matrix factorization with a multi-layer perceptron to capture both linear and non-linear user-item interactions in a unified architecture.

    \item \textbf{NGCF}~\citep{wang2019neural}: Neural Graph Collaborative Filtering explicitly incorporates high-order connectivity in graph message passing, enhancing user and item embeddings with both structural and semantic signals.

    \item \textbf{SimpleX}~\citep{mao2021simplex}: SimpleX adopts a margin-based contrastive objective and soft negative sampling to simplify self-supervised learning, achieving strong performance without complex augmentations.

    \item \textbf{SASRec}~\citep{kangSelfAttentiveSequential2018}: Self-Attention based Sequential Recommendation employs a self-attention mechanism to capture sequential patterns in user behavior, effectively modeling temporal dynamics.

    \item \textbf{NARM}~\citep{Li2017NeuralAtt}: Neural Attentive Session-based Recommendation models user behavior with an attention mechanism, effectively capturing temporal dynamics and user preferences.
\end{itemize}

All model implementations and training configurations followed the publicly available library, RecBole~\citep{recbole[1.0]}. 
The specific hyperparameter settings are detailed below:
\begin{table}[h]
\centering
\scriptsize
\label{tab:hyperparams}
\begin{tabular}{ll}
\toprule
\textbf{Model} & \textbf{Hyperparameters} \\
\midrule
\textbf{BPR} & Embedding dimension = 64; Learning rate = 0.001 \\
\midrule
\textbf{LightGCN} & Embedding dimension = 64; \#Layers = 2; \\
                  & Learning rate = 0.001; $L_2$ regularization = 0.001 \\
\midrule
\textbf{MultiDAE} & Latent dimension = 64; Hidden layer size = [600]; \\
                  & Learning rate = 0.001; Dropout rate = 0.5 \\
\midrule
\textbf{DiffRec} & Noise schedule = linear; Noise scale = 0.001; \\
                 & Noise range = [0.0005, 0.005]; DNN layer size = [300]; \\
                 & Learning rate = 0.0001; $L_2$ regularization = $10^{-5}$; \\
                 & Diffusion steps = 5; $w_\text{min}$ = 0.1 \\
\midrule
\textbf{SGL} & Graph type = ED; \#Layers = 2; $\tau$ = 0.5; \\
             & Embedding dim = 64; Edge dropout = 0.1; \\
             & Learning rate = 0.001; $L_2$ = $10^{-5}$; SSL weight = 0.05 \\
\midrule
\textbf{NeuMF} & MF dim = 64; MLP dim = 64; MLP layers = [64, 32, 16]; \\
               & Dropout rate = 0.1; Learning rate = 0.001; \\
\midrule
\textbf{NGCF} & Embedding dim = 64; Message dropout = 0.1; \\
              & $L_2$ weight = $10^{-5}$; \#Layers = 3 \\
\midrule
\textbf{SimpleX} & Embedding dim = 64; Margin = 0.9; Neg weight = 10; \\
                 & $L_2$ weight = $10^{-5}$; Gamma = 0.5 \\
\midrule
\textbf{SASRec} & Embedding dim = 64; Inner size = 256; Num layers = 2; \\
                 & Learning rate = 0.001; Dropout rate = 0.5 \\
\midrule
\textbf{NARM} & Embedding dim = 64; Hidden size = 128; Num layers = 1; \\
                 & Learning rate = 0.001; Dropout rate = 0.5 \\
\bottomrule
\end{tabular}
\caption{Hyperparameter configurations for backbone recommendation models.}
\end{table}

\textbf{Step 2: Simulating User Behaviors.}
We utilized each trained backbone recommendation model to generate a candidate exposure list of length $10$ for each user. 
Based on this list, we simulated user behaviors using two models: a base LLM (LLM Base) and a fine-tuned user simulator (SFT+DPO). 
Each model selects ab item from the exposure list to represent the simulated user–item interaction ($\mathcal{D}_{sim}$). 
In this experiment, we use \textit{Llama-3.2-3B-Instruct} as the base LLM.

\textbf{Step 3: Retraining with Simulated Interactions.}
Step 3: Retraining with Simulated Interactions.
The simulated user–item interactions obtained in Step 2 are incorporated into the original training set to create an augmented dataset, $\mathcal{D}_{train}^{aug} = \mathcal{D}_{train} \cup \mathcal{D}_{sim}$.
We then retrain the backbone recommendation models using the same procedure outlined in Step 1. 
By comparing model performance before and after the augmentation, we assess the effectiveness of the simulated user behaviors in improving recommendation quality.

\subsection{Detailed Setup of Different Data Selection Strategies}
\label{Append:Ablation_Setup}
In Section \ref{Subsec:DetailedAnalysis}, we compare the proposed uncertainty decomposition based data distillation method with several baselines. Here we list the detailed setup of each baseline.
To ensure a fair comparison, all strategies are adopted on the same candidate sample set we processed in Section 3.2, with only the following changes: 
\begin{itemize}[leftmargin=24pt]
    \item \textbf{Random (w/o Decisions)}: We use random selection to replace the difference of epistemic uncertainties $\Delta(\rmX, (A, B))$ in \eqref{Eq:DiffUncert}, and use the direct behavior prediction as the same as Section 3.1.
    \item \textbf{Random (w/ Decisions)}: We use random selection to replace the difference of epistemic uncertainties $\Delta(\rmX, (A, B))$ in \eqref{Eq:DiffUncert}.
    \item \textbf{High Accuracy}: We select the samples where the powerful LLM \textit{Qwen-2.5-32B-Instruct} achieves the highest average accuracy with the 10 sampling results.
    \item \textbf{Low Accuracy}: We select the samples where the powerful LLM \textit{Qwen-2.5-32B-Instruct} achieves the lowest average accuracy with the 10 sampling results.
    \item \textbf{Diff. Accuracy}: We select the samples where the powerful LLM \textit{Qwen-2.5-32B-Instruct} and weak LLM \textit{Llama-3.2-3B-Instruct} achieve the maximum difference in the average accuracy with the 10 sampling results.
    \item \textbf{IFD Score~\citep{liSuperfilteringWeaktoStrong2024}}: We use the Instruction-Following Difficulty (IFD) score, which measure the difference of the perplexity of the generated decision-making processes with and without the prompt. We use \textit{Llama-3.2-3B-Instruct} to calculate the perplexity.
\end{itemize}

\section{Additional Experiment Results}

\subsection{Interactive Simulation with RS}
In addition to the results of the four primary models presented in Table~\ref{Tab:Rec_Improvement}, the performance of the remaining six models (BPR, MultiDAE, SGL, NeuMF, NGCF and SimpleX) is reported in Table~\ref{Tab:Rec_Improvement_2}. 
These additional results serve to further validate the consistency and robustness of our findings across a broader range of model configurations. 
By including these supplementary models in the appendix, we aim to provide a more comprehensive evaluation and ensure that our conclusions are not limited to a specific subset of experimental settings.
\begin{table*}[ht]
\centering
\renewcommand{\arraystretch}{0.85}
\scriptsize
\setlength\tabcolsep{5pt}
\setul{1pt}{0.6pt}
\begin{tabular}{@{}cc|cccccc|cccccc@{}}
\toprule
\multicolumn{2}{c|}{\multirow{2}{*}{\textbf{Model}}} & \multicolumn{6}{c|}{\textbf{Movielens}} & \multicolumn{6}{c}{\textbf{Steam}} \\ \cmidrule(l){3-14} 
\multicolumn{2}{c|}{} & \textbf{R@5} & \textbf{R@10} & \textbf{N@5} & \textbf{N@10} & \textbf{M@5} & \textbf{M@10} & \textbf{R@5} & \textbf{R@10} & \textbf{N@5} & \textbf{N@10} & \textbf{M@5} & \textbf{M@10} \\ \midrule

\multirow{4}{*}{\textbf{BPR}} & \textbf{Backbone} & 4.970 & 8.403 & 3.138 & 4.239 & 2.541 & 2.990 & 3.268 & 5.308 & 2.004 & 2.660 & 1.593 & 1.861 \\
 & \textbf{Base LLM} & 5.235 & 8.725 & 3.249 & 4.371 & 2.602 & 3.062 & 3.254 & 5.316 & 2.067 & 2.729 & 1.680 & 1.950 \\
 & \textbf{SFT+DPO} & \textbf{5.520} & \textbf{9.256} & \textbf{3.531} & \textbf{4.727} & \textbf{2.883} & \textbf{3.370} & \textbf{3.290} & \textbf{5.381} & \textbf{2.171} & \textbf{2.844} & \textbf{1.805} & \textbf{2.081} \\ \cmidrule(l){2-14} 
 & \textbf{Improve} & 11.1\% & 10.2\% & 12.5\% & 11.5\% & 13.5\% & 12.7\% & 0.7\% & 1.4\% & 8.3\% & 6.9\% & 13.3\% & 11.8\% \\ \midrule
\multirow{4}{*}{\textbf{MultiDAE}} & \textbf{Backbone} & 5.046 & 8.953 & 3.314 & 4.562 & 2.750 & 3.255 & 3.173 & 5.535 & 1.973 & 2.733 & 1.582 & 1.894 \\
 & \textbf{Base LLM} & 5.672 & 9.503 & 3.643 & 4.867 & 2.982 & 3.478 & 3.544 & 5.664 & 2.245 & 3.020 & 1.759 & 2.075 \\
 & \textbf{SFT+DPO} & \textbf{6.392} & \textbf{10.015} & \textbf{4.244} & \textbf{5.410} & \textbf{3.543} & \textbf{4.021} & \textbf{3.678} & \textbf{5.974} & \textbf{2.423} & \textbf{3.160} & \textbf{2.015} & \textbf{2.316} \\ \cmidrule(l){2-14} 
 & \textbf{Improve} & \multicolumn{1}{l}{26.7\%} & \multicolumn{1}{l}{11.9\%} & \multicolumn{1}{l}{28.0\%} & \multicolumn{1}{l}{18.6\%} & \multicolumn{1}{l}{28.9\%} & \multicolumn{1}{l|}{23.5\%} & \multicolumn{1}{l}{15.9\%} & \multicolumn{1}{l}{7.9\%} & \multicolumn{1}{l}{22.8\%} & \multicolumn{1}{l}{15.7\%} & \multicolumn{1}{l}{27.3\%} & \multicolumn{1}{l}{22.3\%} \\ \midrule
\multirow{4}{*}{\textbf{SGL}} & \textbf{Backbone} & 4.780 & 8.384 & 2.981 & 4.139 & 2.397 & 2.870 & 3.415 & 5.762 & 2.143 & 2.898 & 1.729 & 2.039 \\
 & \textbf{Base LLM} & 5.425 & 9.048 & 3.426 & 4.590 & 2.773 & 3.249 & 3.641 & 6.193 & 2.273 & 3.089 & 1.827 & 2.159 \\
 & \textbf{SFT+DPO} & \textbf{5.728} & \textbf{9.674} & \textbf{3.709} & \textbf{4.975} & \textbf{3.050} & \textbf{3.566} & \textbf{3.897} & \textbf{6.105} & \textbf{2.476} & \textbf{3.183} & \textbf{2.012} & \textbf{2.300} \\ \cmidrule(l){2-14} 
 & \textbf{Improve} & 19.8\% & 15.4\% & 24.4\% & 20.2\% & 27.3\% & 24.2\% & 14.1\% & 6.0\% & 15.5\% & 9.8\% & 16.4\% & 12.8\% \\ \midrule
\multirow{4}{*}{\textbf{NeuMF}} & \textbf{Backbone} & 3.699 & 7.398 & 2.297 & 3.472 & 1.839 & 2.312 & 3.254 & 5.440 & 2.047 & 2.745 & 1.654 & 1.937 \\
 & \textbf{Base LLM} & 4.818 & \textbf{8.441} & 3.061 & \textbf{4.213} & 2.482 & \textbf{2.947} & \textbf{3.283} & \textbf{5.754} & 2.025 & \textbf{2.823} & 1.614 & 1.943 \\
 & \textbf{SFT+DPO} & \textbf{4.894} & \textbf{8.043} & \textbf{3.111} & \textbf{4.116} & \textbf{2.530} & \textbf{2.938} & 3.224 & 5.579 & \textbf{2.052} & 2.808 & \textbf{1.670} & \textbf{1.978} \\ \cmidrule(l){2-14} 
 & \textbf{Improve} & 32.3\% & 8.7\% & 35.4\% & 18.5\% & 37.6\% & 27.1\% & -0.9\% & 2.6\% & 0.2\% & 2.3\% & 0.9\% & 2.1\% \\ \midrule
\multirow{4}{*}{\textbf{NGCF}} & \textbf{Backbone} & 5.292 & 8.763 & 3.487 & 4.584 & 2.900 & 3.339 & 3.583 & 5.952 & 2.253 & 3.011 & 1.819 & 2.128 \\
 & \textbf{Base LLM} & 5.634 & 9.427 & 3.596 & 4.802 & 2.933 & 3.419 & 3.656 & 6.142 & 2.267 & 3.069 & 1.813 & 2.143 \\
 & \textbf{SFT+DPO} & \textbf{6.108} & \textbf{9.826} & \textbf{4.024} & \textbf{5.213} & \textbf{3.344} & \textbf{3.827} & \textbf{3.744} & \textbf{6.200} & \textbf{2.360} & \textbf{3.148} & \textbf{1.908} & \textbf{2.230} \\ \cmidrule(l){2-14} 
 & \textbf{Improve} & 15.4\% & 12.1\% & 15.4\% & 13.7\% & 15.3\% & 14.6\% & 4.5\% & 4.2\% & 4.7\% & 4.6\% & 4.9\% & 4.8\% \\ \midrule
\multirow{4}{*}{\textbf{SimpleX}} & \textbf{Backbone} & 4.742 & 8.080 & 2.908 & 3.973 & 2.308 & 2.739 & 3.649 & 6.193 & 2.292 & 3.107 & 1.850 & 2.182 \\
 & \textbf{Base LLM} & 4.989 & 8.062 & 3.081 & 4.065 & 2.463 & 2.863 & 3.817 & \textbf{6.361} & 2.342 & 3.157 & 1.860 & 2.192 \\
 & \textbf{SFT+DPO} & \textbf{5.292} & \textbf{8.801} & \textbf{3.307} & \textbf{4.431} & \textbf{2.655} & \textbf{3.113} & \textbf{4.197} & \textbf{6.895} & \textbf{2.790} & \textbf{3.660} & \textbf{2.330} & \textbf{2.687} \\ \cmidrule(l){2-14} 
 & \textbf{Improve} & 11.6\% & 8.9\% & 13.7\% & 11.5\% & 15.0\% & 13.6\% & 15.0\% & 11.3\% & 21.7\% & 17.8\% & 25.9\% & 23.2\% \\ \bottomrule
\end{tabular}
\caption{Performance comparison of user simulators on recommendation scenarios (Cont.). (In percentage)}
\label{Tab:Rec_Improvement_2}
\end{table*}

\subsection{Factors}
\label{Append:Factors}
\begin{figure*}[htbp]
    \centering
    \begin{subfigure}[t]{0.32\textwidth}
        \centering
        \includegraphics[width=\linewidth]{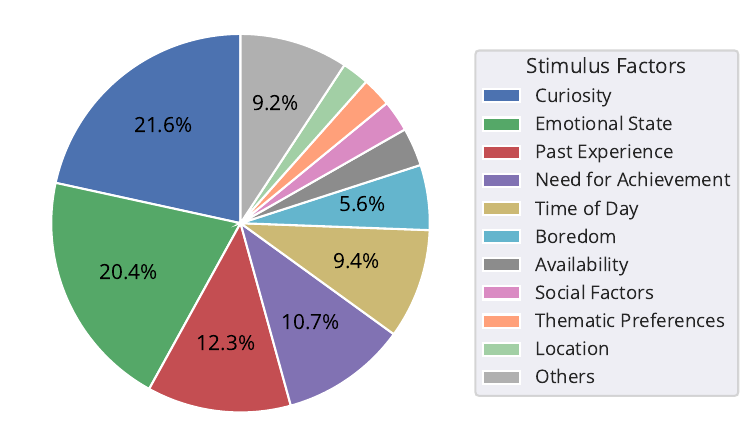}
        \caption{Knowledge Factors}
        \label{fig:stimulus}
    \end{subfigure}
    \hfill
    \begin{subfigure}[t]{0.32\textwidth}
        \centering
        \includegraphics[width=\linewidth]{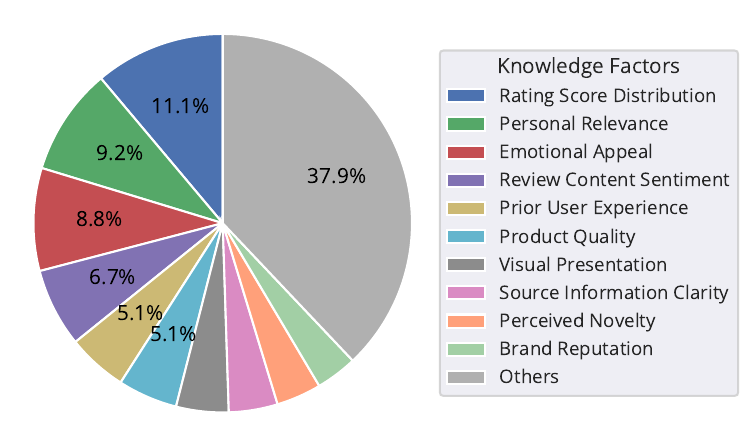}
        \caption{Stimulus Factors}
        \label{fig:knowledge}
    \end{subfigure}
    \hfill
    \begin{subfigure}[t]{0.32\textwidth}
        \centering
        \includegraphics[width=\linewidth]{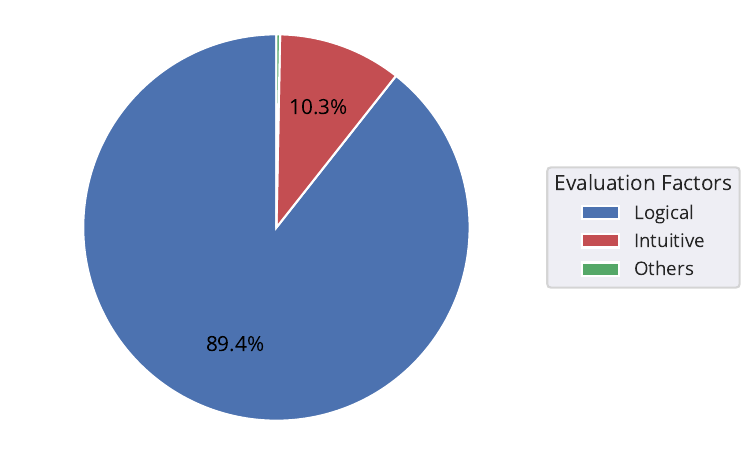}
        \caption{Evaluation Factors}
        \label{fig:cognitive}
    \end{subfigure}

    \caption{An overview of the three factor categories: knowledge, stimulus, and evaluation factors.}
    \label{fig:factor_piecharts}
\end{figure*}
To further understand the reasoning patterns exhibited by the fine-tuned LLM during user simulation, we conducted a statistical analysis of the factors referenced by the model when processing recommendation tasks. 
As illustrated in Figure~\ref{fig:stimulus}, stimulus-related factors are dominated by Curiosity \textit{(21.6\%)} and Emotional State \textit{(20.4\%)}, suggesting that the model frequently simulates users as being driven by intrinsic motivations and affective conditions. 
Other notable factors include Past Experience \textit{(12.3\%)}, Need for Achievement \textit{(10.7\%)}, and Time of Day \textit{(9.4\%)}, indicating the model’s sensitivity to both psychological and contextual signals. 
The remaining factors, such as Boredom \textit{(5.6\%)}, Availability \textit{(3.3\%)}, Social Factors \textit{(2.7\%)}, Thematic Preferences \textit{(2.5\%)}, and Location \textit{(2.3\%)}, contribute to a more nuanced understanding of user context, while Others account for 9.2\%.

Regarding knowledge-based reasoning, the LLM most frequently relies on Rating Score Distribution \textit{(11.1\%)}, Personal Relevance \textit{(9.2\%)}, and Emotional Appeal \textit{(8.8\%)}. 
These factors indicate a hybrid reasoning pattern, where both objective signals and affective resonance shape user preferences. 
Additional sources such as Review Content Sentiment \textit{(6.7\%)}, Prior User Experience \textit{(5.1\%)}, and Product Quality \textit{(5.1\%)} further inform the model’s decisions. 
Less emphasized yet still relevant are factors like Visual Presentation \textit{(4.5\%)}, Source Information Clarity \textit{(4.2\%)}, Perceived Novelty \textit{(3.8\%)}, and Brand Reputation \textit{(3.5\%)}. 
A significant portion \textit{(37.9\%)} is grouped under Others, suggesting a broad and diverse interpretation of user-relevant knowledge by the model.

In terms of evaluation strategy, the LLM overwhelmingly adopts a Logical reasoning approach \textit{(89.4\%)}, with relatively fewer instances of Intuitive decisions \textit{(10.3\%)} and minimal reliance on Others \textit{(0.3\%)}. 
This highlights the model’s preference for rational and structured decision-making when simulating user behavior, aligning with deliberative human reasoning processes.

\subsection{Feature Ablation Study}
\begin{table*}[ht]
\renewcommand{\arraystretch}{1.2}\
\setlength\tabcolsep{2pt}
\centering
\footnotesize
\begin{tabular}{lcc}
\toprule
\textbf{Feature} & \textbf{Accuracy (Retained)} & \textbf{Accuracy (Removed)} \\
\midrule
History         & 0.2578 & \textbf{0.7188} \\
Price           & 0.3135 & 0.7041 \\
Player Reviews  & 0.3770 & 0.7002 \\
Specifications  & 0.3672 & 0.6807 \\
Tags            & 0.5059 & 0.6445 \\
Released Time   & 0.3389 & 0.6953 \\
Developer       & 0.3203 & 0.6973 \\
Title           & \textbf{0.5400} & 0.5938 \\
\bottomrule
\end{tabular}
\caption{Accuracy on the Steam dataset under feature ablation settings}
\label{tab:steam_feature_ablation}
\end{table*}

\begin{table*}[ht]
\renewcommand{\arraystretch}{1.2}\
\setlength\tabcolsep{2pt}
\centering
\footnotesize
\begin{tabular}{lcc}
\toprule
\textbf{Feature} & \textbf{Accuracy (Retained)} & \textbf{Accuracy (Removed)} \\
\midrule
Rating                     & 0.4658 & 0.8291 \\
History                    & 0.2637 & 0.8652 \\
Published at (Interaction Time) & 0.6865 & 0.8535 \\
Author                     & 0.6836 & 0.8506 \\
Description                & 0.2656 & 0.8633 \\
Tags                       & 0.2510 & \textbf{0.8730} \\
Title                      & \textbf{0.7080} & 0.8291 \\
\bottomrule
\end{tabular}
\caption{Accuracy on the Goodreads aataset under feature ablation settings.}
\label{tab:goodreads_feature_ablation}
\end{table*}

Table~\ref{tab:steam_feature_ablation} and Table~\ref{tab:goodreads_feature_ablation} present feature ablation results on two distinct domains: the Steam dataset (video games) and the Goodreads dataset (books). Each table reports model accuracy when either only a single feature is retained or when a single feature is removed, highlighting the independent predictive power and marginal contribution of features, respectively.

For the Steam dataset, \textit{Title} and \textit{Tags} stand out as the most informative features both individually and collectively, with Title removal causing the largest accuracy drop. 
This reflects the crucial role of semantic identifiers and categorical tags in characterizing games, which aligns well with the strong textual and categorical priors that large language models (LLMs) possess about gaming content. 
Conversely, features such as \textit{History} and \textit{Developer} contribute less individually and appear somewhat redundant when other features are present, possibly due to overlapping information captured by \textit{Tags} and \textit{Player Reviews}.

In contrast, for the Goodreads dataset, features tied to content metadata and contextual information—namely \textit{Title}, Published time, and \textit{Author}—exhibit the highest individual predictive power. 
Their removal significantly impacts performance, underscoring the importance of temporal and authorial context in book recommendation tasks. 
Features like \textit{History}, \textit{Description}, and \textit{Tags} show relatively weaker standalone influence and limited marginal contribution, suggesting that, in literary domains, user interaction history and descriptive tags may be less decisive compared to core bibliographic information.

These domain-specific patterns reflect differences in the nature of the datasets and the prior knowledge encoded within LLMs. 
In gaming, categorical labels and titles are rich semantic anchors, supporting detailed user preference modeling. 
Meanwhile, in books, temporal context and authorship carry substantial weight, likely because publication timing and author identity strongly shape readers' interests and reception. 
LLMs trained on broad textual corpora inherently capture such domain-dependent nuances, which is reflected in the differential feature importance observed.

Overall, this comparative study highlights that while \textit{Title} is universally critical across domains, the secondary features that most effectively support recommendations differ—\textit{Tags} and \textit{Specifications} for games versus \textit{Published time} and \textit{Author} for books—emphasizing the need for domain-aware feature selection and modeling strategies when leveraging LLMs for recommendation tasks.

\subsection{Analysis of Confidence Level Calibration}
To examine how training affects model confidence, we analyze the entropy of behavior prediction $\mathcal{H}(P(\bm Y | \bm X \oplus \bm C))$ in \eqref{eq:our_decompose} following the setting in Section \ref{Subsec:Data_Distillation} on the test set across different training stages.
\begin{table}[ht]
\footnotesize
\renewcommand{\arraystretch}{1.2}
\setlength\tabcolsep{6pt}
\centering
\begin{tabular}{lc}
\toprule
\multicolumn{1}{c}{\textbf{Model}} & \multicolumn{1}{c}{\textbf{Entropy}} \\
\midrule
Base (Llama-3.2-3B-Instruct) & $1.252 \pm 0.138$ \\
\:\:+ SFT & $0.219 \pm 0.064$ \\
\:\:+ SFT + DPO & $0.074 \pm 0.032$ \\
\:\:+ SFT + GRPO & $0.019 \pm 0.009$ \\ \midrule
Teacher (Qwen-2.5-32B-Instruct) & $0.011 \pm 0.006$ \\
\bottomrule
\end{tabular}
\caption{Entropy of behavior prediction across different training stages on the test set.}
\label{tab:confidence_entropy}
\end{table}

Table~\ref{tab:confidence_entropy} presents the entropy of behavior predictions across different training stages. 
The results reveal a substantial and progressive reduction in prediction entropy as the model undergoes supervised fine-tuning (SFT) and preference optimization (DPO/GRPO).
The base model (Llama-3.2-3B-Instruct) exhibits relatively high entropy, indicating considerable uncertainty in its behavior predictions. 
After SFT, entropy drops dramatically, demonstrating that supervised learning on high-quality decision-making processes significantly improves the model's confidence. 
Further refinement through DPO reduces entropy, and GRPO achieves even lower entropy, approaching the confidence level of the teacher model (Qwen-2.5-32B-Instruct).

The observed entropy reduction through training stages is both expected and desirable for a user simulator generating complete decision-making processes. 
Lower entropy indicates that the model produces less ambiguous evaluations and more definitive, high-confidence choices, which aligns with the goal of simulating realistic user behavior where users typically make clear decisions after deliberation. 
The progressive confidence increase from base model to SFT to preference optimization stages demonstrates that our training pipeline successfully guides the model toward more decisive and human-like decision-making patterns.

\subsection{Style Analysis of Decision-Making Processes}
To better understand the decision-making processes of the user simulators, we conduct a statistical analysis of the generated decision-making process across different training stages on the test set.
We measure several metrics to characterize the decision-making processes, including the length of generations and automated text similarity metrics (BLEU and ROUGE) using the teacher model's decision processes as reference.
\begin{table}[ht]
\scriptsize
\renewcommand{\arraystretch}{1.2}
\setlength\tabcolsep{4pt}
\centering
\begin{tabular}{lcccc}
\toprule
\textbf{Model} & \textbf{ROUGE-L} & \textbf{ROUGE-2} & \textbf{BLEU-2} & \textbf{Length} \\
\midrule
Llama-3.2-3B-Instruct & 0.300 & 0.239 & 0.186 & 273 \\
\:\:+ SFT & 0.377 & 0.291 & 0.264 & 218 \\
\:\:+ SFT + GRPO & 0.376 & 0.287 & 0.261 & 217 \\
\:\:+ SFT + DPO & 0.376 & 0.307 & 0.266 & 238 \\
\midrule
Qwen2.5-3B-Instruct & 0.282 & 0.176 & 0.159 & 179 \\
\:\:+ SFT & 0.380 & 0.293 & 0.264 & 220 \\
\:\:+ SFT + GRPO & 0.380 & 0.292 & 0.264 & 220 \\
\:\:+ SFT + DPO & 0.375 & 0.313 & 0.263 & 253 \\
\midrule
Qwen2.5-32B-Instruct &   &   &   & 218 \\
\bottomrule
\end{tabular}
\caption{Statistical analysis of decision-making process styles across training stages on the test set. ROUGE and BLEU scores are computed using the teacher model's decision processes as reference.}
\label{tab:reasoning_style_analysis}
\end{table}

Table~\ref{tab:reasoning_style_analysis} presents the statistical analysis of reasoning styles across different training stages.
Several key insights emerge from these results:

\textbf{Consistent reasoning length.} The token length of generated decision-making processes remains relatively consistent across fine-tuned models and the teacher model, indicating that student models do not drastically shorten or extend their generations after training. This suggests that the models learn to generate appropriately detailed decision processes.

\textbf{SFT learns reasoning structure.} Supervised fine-tuning (SFT) dramatically increases ROUGE and BLEU scores compared to base models, demonstrating that students primarily learn the teacher's reasoning style and structural patterns during the SFT phase. This indicates successful knowledge transfer of the decision-making framework.

\textbf{Preference optimization refines content.} The DPO and GRPO stages show largely unchanged text similarity metrics despite significantly improved user preference alignment. This suggests that preference optimization primarily refines the semantic content and decision quality within the already learned reasoning style acquired during SFT, rather than altering the structural patterns of the decision-making process.

These findings reveal that our training pipeline effectively transfers both the reasoning structure (via SFT) and decision quality (via preference optimization) from the teacher model to student models, resulting in user simulators that generate human-like decision-making processes with high confidence and alignment with actual user preferences.

\subsection{Domain Generalization of User Simulators}
\label{subsec:domain_generalization}
To evaluate the generalizability of our trained user simulator across different domains, we conduct a specialized cross-domain experiment. We split our multi-domain dataset into two disjoint groups and train models on one group while testing on the other, simulating a zero-shot cross-domain transfer scenario.

\textbf{Experimental Setup.} We partition the datasets as follows: \textbf{Group 1}: \textit{Goodreads, MobileRec, MIND, Amazon (Fashion, Office)} and \textbf{Group 2}: \textit{KuaiRec2, LastFM, MovieLens, Steam, Amazon (Grocery, Beauty)}.
Then, we train our model using SFT+DPO on each group separately and evaluate its performance on the held-out group, comparing against the base model, the full-data trained model, and GPT-5 as the baseline.

\begin{table*}[ht]
\scriptsize
\renewcommand{\arraystretch}{1.0}
\setlength\tabcolsep{3pt}
\centering
\begin{tabular}{l|ccccc|cccccc}
\toprule
\multirow{2}{*}{\textbf{Model}} & \multicolumn{5}{c|}{\textbf{Group 1 (Trained on Group 2)}} & \multicolumn{6}{c}{\textbf{Group 2 (Trained on Group 1)}} \\
\cmidrule(lr){2-6} \cmidrule(lr){7-12}
 & \textbf{Fashion} & \textbf{Office} & \textbf{MIND} & \textbf{Goodreads} & \textbf{MobileRec} & \textbf{Beauty} & \textbf{Grocery} & \textbf{KuaiRec2} & \textbf{LastFM} & \textbf{MovieLens} & \textbf{Steam} \\
\midrule
Llama-3.2-3B-Instruct & 22.3 & 21.1 & 19.9 & 27.1 & 22.1 & 21.9 & 22.3 & 22.5 & 18.3 & 23.3 & 26.2 \\
\:\:+ SFT + DPO (Full Data) & 53.9 & 52.6 & 34.0 & 86.9 & 34.7 & 49.8 & 51.0 & 48.7 & 35.7 & 67.3 & 70.0 \\
\:\:+ SFT + DPO (Cross-Domain) & 48.8 & 41.2 & 26.1 & 68.5 & 31.4 & 37.4 & 45.8 & 39.2 & 28.1 & 54.6 & 50.0 \\
\midrule
GPT-5 & 46.3 & 35.0 & 33.2 & 78.9 & 25.3 & 41.6 & 38.1 & 30.8 & 37.2 & 61.1 & 45.2 \\
\bottomrule
\end{tabular}
\caption{Cross-domain generalization performance. Models trained on one group are evaluated on the other group in a zero-shot transfer setting. Accuracy (\%) is reported for all domains.}
\label{tab:cross_domain_generalization}
\end{table*}

Table~\ref{tab:cross_domain_generalization} presents the cross-domain generalization results. The cross-domain fine-tuned models demonstrate substantial performance improvements over the base model across nearly all domains. This indicates that the decision-making patterns learned from one set of domains effectively transfer to unseen domains.
Notably, in several domains (e.g., Amazon Fashion, Amazon Office, Amazon Grocery, Goodreads), the cross-domain fine-tuned model outperforms GPT-5. This demonstrates that the learned decision-making logic possesses meaningful domain generalization capability and is not merely memorizing domain-specific patterns.
While cross-domain performance is strong, it generally falls short of the full-data trained model. This gap highlights the value of our multi-domain data construction approach, which enables the model to capture a more comprehensive and shared understanding of user decision-making across diverse contexts.
The performance gap between cross-domain and full-data training varies across domains. For instance, MIND and LastFM show larger gaps, suggesting these domains may have more distinctive user behavior patterns that benefit from in-domain training data.

These findings validate that our framework learns generalizable decision-making patterns rather than domain-specific heuristics, while also confirming the necessity of multi-domain training for optimal performance across diverse recommendation scenarios.

\section{Prompts and Templates}
\subsection{Prompt Templates of User Simulation Scene Construction}
\label{subsec:appendix_prompts}
We designed prompts based on user feedback datasets to construct conversational recommendation scenes. 
The prompts mainly consist of three components: first, the \textbf{user profile}, which introduces the user's personal information; second, the \textbf{interaction history}, which outlines the user-item interaction; and third, exposures with the description of each exposed items.
Upon the available information in each domain, we provide a detailed template for datasets of each domain as follows.

\begin{tcolorbox}[colframe=black!65!white, colback=black!5!white, fontupper=\slshape, title=Template of Movielens]
\%\textbf{Template of user profile}
\scriptsize
\#\# User Profile \\
Gender: \highlight{\{GENDER\}} \\
Age: \highlight{\{AGE\}} \\
Occupation: \highlight{\{OCCUPATION\}} \\
Location: \highlight{\{LOCATION\}} \\

\%\textbf{Template of interaction history}

A movie viewed \highlight{\{TIME DIFF\}} \\
\highlight{\{TITLE\}} - \highlight{\{GENRES\}} \\
Rating: \highlight{\{RATING\}}/5.0

\%\textbf{Template of exposure list}

\highlight{NEW RELEASE}\highlight{\{TITLE\}} - \highlight{\{GENRES\}} \\

\end{tcolorbox}

\begin{tcolorbox}[colframe=black!65!white, colback=black!5!white, fontupper=\slshape, title=Template of Amazon]
    \scriptsize
    \%\textbf{Template of user profile}
    
    \#\# User Profile \\
    
    \%\textbf{Template of interaction history}
    
    A product bought \highlight{\{TIME DIFF\}} \\
    Product name: \highlight{\{SHORT TITLE\}} \\
    Category: \highlight{\{CATEGORY\}} \\
    Price: \highlight{\{REV PRICE\}} \\
    Ratings: \highlight{\{AVERAGE RATING\}} (\highlight{\{RATING NUMBER\}}) \\
    My Behavior: \\
    Rating: \highlight{\{RATING\}}/5 \\
    Review: \highlight{\{REVIEW SUMMARY\}} \\
    \highlight{\{REVIEW TEXT\}} \\
    
    \%\textbf{Template of exposure list}
    
    Product name: \highlight{\{SHORT TITLE\}} \\
    Category: \highlight{\{CATEGORY\}} \\
    Price: \highlight{\{REV PRICE\}} \\
    Ratings: \highlight{\{AVERAGE RATING\}} (\highlight{\{RATING NUMBER\}}) \\

    \end{tcolorbox}

\begin{tcolorbox}[colframe=black!65!white, colback=black!5!white, fontupper=\slshape, title=Template of KuaiRec]
    \scriptsize
    \%\textbf{Template of user profile}
    
    \#\# User Profile \\
    Country: China \\
    Registered \highlight{\{REGISTER DAYS\}} days ago \\
    \highlight{\{FRIEND USER NUM\}} Friends - \highlight{\{FOLLOW USER NUM\}} Followers - \highlight{\{FANS USER NUM\}} Fans \\
    Active Degree: \highlight{\{ACTIVE DEGREE\}} \\
    Role: \highlight{\{ROLE\}} \\
    
    \%\textbf{Template of interaction history}
    
    A video watched \highlight{\{TIME DIFF\}} \\
    Title: \highlight{\{CAPTION\}} \highlight{\{CATEGORY\}} \highlight{\{COVER TEXT\}} \\
    Duration: \highlight{\{VIDEO DURATION\}} \\
    \highlight{\{PLAY COUNT\}} plays - \highlight{\{LIKE COUNT\}} likes - \highlight{\{COMMENT COUNT\}} comments - \highlight{\{SHARE COUNT\}} shares \\
    My Behavior: \highlight{\{OPERATION\}} \\
    
    \%\textbf{Template of exposure list}
    
    Title: \highlight{\{CAPTION\}} \highlight{\{CATEGORY\}} \highlight{\{COVER TEXT\}} \\
    Duration: \highlight{\{VIDEO DURATION\}} \\
    \highlight{\{PLAY COUNT\}} plays - \highlight{\{LIKE COUNT\}} likes - \highlight{\{COMMENT COUNT\}} comments - \highlight{\{SHARE COUNT\}} shares \\

    \end{tcolorbox}

\begin{tcolorbox}[colframe=black!65!white, colback=black!5!white, fontupper=\slshape, title=Template of Goodreads]
\scriptsize
\%\textbf{Template of user profile}

\#\# User Profile \\

\%\textbf{Template of interaction history}

A book viewed \highlight{\{TIME DIFF\}} \\
Title: \highlight{\{TITLE\}} (\highlight{\{GENRES\}}) \\
Description: \highlight{\{DESCRIPTION\}} \\
Author: \highlight{\{AUTHORS\}} \\
Published at \highlight{\{PUBLICATION YEAR\}} - \highlight{\{PUBLISHER\}} - \highlight{\{NUM PAGES\}} pages \\
Rating: \highlight{\{AVERAGE RATING\}} - \highlight{\{RATINGS COUNT\}} ratings \\
My Behavior: \highlight{\{READ STATUS\}} \highlight{\{RATING UPDATED\}} \highlight{\{REVIEW UPDATED\}} \\

\%\textbf{Template of exposure list}

Title: \highlight{\{TITLE\}} (\highlight{\{GENRES\}}) \\
Author: \highlight{\{AUTHORS\}} \\
Published at \highlight{\{PUBLICATION YEAR\}} - \highlight{\{PUBLISHER\}} - \highlight{\{NUM PAGES\}} pages \\
Rating: \highlight{\{AVERAGE RATING\}} - \highlight{\{RATINGS COUNT\}} ratings \\

\end{tcolorbox}

\begin{tcolorbox}[colframe=black!65!white, colback=black!5!white, fontupper=\slshape, title=Template of MobileRec]
\scriptsize
\%\textbf{Template of user profile}

\#\# User Profile \\

\%\textbf{Template of interaction history}

\highlight{\{APP NAME\}} (\highlight{\{AVG RATING\}}/5.0 - \highlight{\{NUM REVIEWS\}} reviews) \\
Category: \highlight{\{APP CATEGORY\}} \\
Developer: \highlight{\{DEVELOPER NAME\}} \\
Price: \highlight{\{PRICE\}} \\
Description: \highlight{\{SHORT DESCRIPTION\}} \\
My Rating: \highlight{\{RATING\}} \\
My Reviews: \highlight{\{REVIEW\}} \\

\%\textbf{Template of exposure list}

\highlight{\{APP NAME\}} (\highlight{\{AVG RATING\}}/5.0 - \highlight{\{NUM REVIEWS\}} reviews) \\
Category: \highlight{\{APP CATEGORY\}} \\
Developer: \highlight{\{DEVELOPER NAME\}} \\
Price: \highlight{\{PRICE\}} \\

\end{tcolorbox}

\begin{tcolorbox}[colframe=black!65!white, colback=black!5!white, fontupper=\slshape, title=Template of LastFM]
\scriptsize
\%\textbf{Template of user profile}

\#\# User Profile \\

\%\textbf{Template of interaction history}

An artist listened \highlight{\{TIME DIFF\}} \\
Artist name: \highlight{\{NAME\}} \\
Tags: \highlight{\{TAGS\}} \\
My Behavior: Tagging with \highlight{\{TAG VALUE\}} \\

\%\textbf{Template of exposure list}

Artist name: \highlight{\{NAME\}} (\highlight{\{LISTENED FRIENDS\}}) \\
Tags: \highlight{\{TAGS\}} \\

\end{tcolorbox}

\begin{tcolorbox}[colframe=black!65!white, colback=black!5!white, fontupper=\slshape, title=Template of Steam]
\scriptsize
\%\textbf{Template of user profile}

\#\# User Profile \\
A user who has purchased \highlight{\{PRODUCTS\}} games \\

\%\textbf{Template of interaction history}

A game purchased \highlight{\{TIME DIFF\}} \\
Title: \highlight{\{APP NAME\}} [\highlight{\{GENRES\}}] \\
Developed by \highlight{\{DEVELOPER\}}  - Published by \highlight{\{PUBLISHER\}} \\
Released: \highlight{\{RELEASE DATE\}} \\
Tags: \highlight{\{TAGS\}} \\
Specifications: \highlight{\{SPECS\}} \\
Player Reviews: \highlight{\{SENTIMENT\}} \\
Price: \highlight{\{PRICE\}} \\
My Behavior: Played \highlight{\{HOURS PLAYED\}} hours  - Review: \highlight{\{REVIEW TEXT\}} \\

\%\textbf{Template of exposure list}

Title: \highlight{\{APP NAME\}} [\highlight{\{GENRES\}}] \\
Developed by \highlight{\{DEVELOPER\}} - Published by \highlight{\{PUBLISHER\}} \\
Released: \highlight{\{RELEASE DATE\}} \\
Tags: \highlight{\{TAGS\}} \\
Specifications: \highlight{\{SPECS\}} \\
Player Reviews: \highlight{\{SENTIMENT\}} \\
Price: \highlight{\{PRICE\}} \\

\end{tcolorbox}

\begin{tcolorbox}[colframe=black!65!white, colback=black!5!white, fontupper=\slshape, title=Template of MIND]
    \%\textbf{Template of user profile}
    \scriptsize
    \#\# User Profile \\
    
    \%\textbf{Template of exposure list}
    
    \highlight{\{TITLE\}} (Category: \highlight{\{CATEGORY\}}) \\
    
    \%\textbf{Template of interaction history}
    
    \highlight{\{TITLE\}} (Category: \highlight{\{CATEGORY\}}) \{ Viewed \highlight{\{TIME DIFF\}} \} \\
\end{tcolorbox}

The text within the curly braces $\{ \}$ represents placeholders that need to be filled with specific information from the dataset. Specifically, the "TIME DIFF" field denotes the time difference between the interaction and the current time, where interactions less than 1 minute old are marked as "just now," and older interactions show "X min/hour/day/month/year ago." It is important to note that different datasets contain varying levels of information, and some datasets may be missing certain components, such as the user profile or user behavior description, where we leave them blank.

Here we show an example of a constructed user simulation scene.

\begin{tcolorbox}[colback=myblue!5!white, colframe=myblue!95!white, fontupper=\slshape, title=Example of the User Simulation Scene]
    \scriptsize
    \begin{multicols}{2}
    
    \textbf{User Profile:} \\
    Gender: \highlight{female} \\
    Age: \highlight{35-44}  \\
    Occupation: \highlight{customer service}  \\
    Location: \highlight{Schenectady, NY}  \\
    
    \textbf{User Interaction History:} 
    
    A movie viewed \highlight{4 minutes ago}: \\
    \quad \highlight{Simple Plan, A (1998) - Crime -Thriller} \\
    \quad Rating: \highlight{4.0}/5.0 \\
    
    A movie viewed \highlight{3 minutes ago}: \\
    \quad \highlight{Crimson Tide (1995) - Drama -Thriller -War} \\
    \quad Rating: \highlight{2.0}/5.0 \\
    
    A movie viewed \highlight{3 minutes ago}: \\
    \quad \highlight{Enemy of the State (1998) - Action -Thriller} \\
    \quad Rating: \highlight{3.0}/5.0 \\

    \columnbreak
    
    A movie viewed \highlight{3 minutes ago}: \\
    \quad \highlight{Die Hard 2 (1990) - Action -Thriller} \\
    \quad Rating: \highlight{3.0}/5.0 \\
    
    A movie viewed \highlight{3 minutes ago}: \\
    \quad \highlight{Six Days Seven Nights (1998) - Adventure -Comedy -Romance} \\
    \quad Rating: \highlight{5.0}/5.0 \\

    \textbf{Exposure List:} 
    
    [A] \highlight{American Beauty (1999) - Comedy  - Drama}
    
    [B] \highlight{[New Release] Hamlet (2000) - Drama}
    
    [C] \highlight{Spawn (1997) - Action  - Adventure  - Sci-Fi  - Thriller}
    
    [D] \highlight{Hana-bi (1997) - Comedy  - Crime  - Drama}
    
    [E] \highlight{Awakenings (1990) - Drama}
    
    [F] \highlight{Drunks (1997) - Drama}
    
    [G] \highlight{Gandhi (1982) - Drama}
    
    [H] \highlight{The Governess (1998) - Drama  - Romance}
    
    [I] \highlight{Whatever (1998) - Drama}
    
    [J] \highlight{Firelight (1997) - Drama}
    
    [K] \highlight{The Best Man (1999) - Drama}
    
    [L] \highlight{My Favorite Year (1982) - Comedy}
    \end{multicols}
    \end{tcolorbox}

 




\subsection{Instruction Prompts for User Simulation Scene}
We design two types of prompts to guide the LLM: (1) direct prediction, and (2) prediction with decision-making process.
The direct prediction prompt directly asks it to predict the user's next action based on the user simulation scene. And the prompt with decision-making process introduces an intermediate reasoning stage inspired by the cognitive process of human decision-making.

\begin{tcolorbox}[colback=myred!5!white,colframe=myred!95!white, fontupper=\slshape, title=Prompt for Direct Prediction]
    \scriptsize
You are a sophisticated user behavior emulator. 
Given a user profile and an exposure list, generate a user behavior in the given exposure list. 
Each choice in the exposure list is indicated by a alphabetic identifier [A], [B], [C], etc. Your output should be a choice identifier from the exposure list, for example, "Behavior: [G]".

\highlight{\{User Profile\}}

\highlight{\{Interaction History\}}

\highlight{\{Exposures\}}

\end{tcolorbox}

\begin{tcolorbox}[colback=myred!5!white,colframe=myred!95!white, fontupper=\slshape, title=Prompt for Behavior Prediction with Decision-making Process]
    \scriptsize
You are a sophisticated user behavior emulator, tasked with simulating user responses within a general recommendation context. Given a user profile and an exposure list, generate a detailed, first-person intent statement that reflects the user's behavior. Your simulations should be adapted for diverse recommendation domains such as media, businesses, and e-commerce. \\

\textbf{Intent Structure and Content:} \\
The intent should be structured as a logical progression through the following stages, each marked by a corresponding label: 
\begin{itemize}[leftmargin=12pt] 
    \item \textbf{Stimulus:} [Describe the initial motivation or need that initiates the user's thought process. This should connect to their profile's spatial, temporal, thematic preferences, causal, and social factors.].
    \begin{itemize}
        \item \textbf{Stimulus Factors:} [List 1-3 most relevant factors from: Internal States (Boredom, Hunger, Thirst, Fatigue/Restlessness, Emotional State, Curiosity, Need for Achievement, Inspiration), External Cues (Time of Day, Day of Week, Weather, Location, Social Factors, Special Occasion, Notification, Advertising, Financial Situation, Availability)].
    \end{itemize}
    \item \textbf{Knowledge:} [Describe the user's thought process as they gain knowledge from the exposure list.  Highlight specific attributes of the options that resonate with the user's preferences, drawing on the user profile.].
    \begin{itemize}
        \item \textbf{Knowledge Factors:} [List 2-4 most influential factors from: Product/Service Attributes (Price, Quality, Features, Convenience, Novelty, Brand Reputation, Personal Relevance (Functional, Thematic, Identity-Based), Emotional Appeal, Time Commitment, Risk), Information Source \& Presentation (Visual Presentation, Recommendation Source, Review Content/Sentiment, Rating Score/Distribution, Social Proof), User's Prior Knowledge (Past Experience, User Preferences/History)].
    \end{itemize}
    \item \textbf{Evaluation:} [Explain the user's internal justification for their preference.] 
    \begin{itemize}
        \item \textbf{Evaluation Style:} [Specify 1 style of the evaluation process, such as Logical, Intuitive, Impulsive, Habitual].
    \end{itemize}

\textbf{Output Format:}

    \textbf{Thought}:

        \setlength{\parindent}{6pt} -Stimulus: [Stimulus Description]

        \setlength{\parindent}{6pt} -Stimulus Factors: [Factor 1], [Factor 2]

        \setlength{\parindent}{6pt} -Knowledge: [Knowledge Description]

        \setlength{\parindent}{6pt} -Knowledge Factors: [Factor 1], [Factor 2], [Factor 3]

        \setlength{\parindent}{6pt} -Evaluation: [Evaluation Description]

        \setlength{\parindent}{6pt} -Evaluation Style: [Evaluation Style]

    \textbf{Behavior}: [Behavior] \\
\end{itemize}

\textbf{Constraints:} 
\begin{itemize}[leftmargin=12pt] 
    \item While multiple behaviors might be considered in the early stages, the final intent and decision should align with a \textbf{single} behavior.
    \item The behavior can be represented by a single label from the choices in the exposure list, enclosed in square brackets (e.g., [X]).
    \item Use ``I" to reflect the first-person perspective of the user.
\end{itemize}

\highlight{\{User Profile\}}

\highlight{\{Interaction History\}}

\highlight{\{Exposures\}}

\end{tcolorbox}

\section{Case Study}
\label{Append:Case}
In the following, we present specific case studies to illustrate how a fine-tuned LLM can effectively perform user simulation tasks. 
Below shows the model input, which includes the task-specific prompt (see Appendix \ref{subsec:appendix_prompts}), the user's profile and interaction history, as well as an exposure list of available options.

\begin{tcolorbox}[colback=myblue!5!white, colframe=myblue!95!white, fontupper=\slshape, title=Example of User Simulation Scene]
    \scriptsize
    You are a sophisticated user behavior emulator... (System message, omitted. See Appendix \ref{subsec:appendix_prompts}) \\
    
    \# \textbf{User Profile}  \\
            Gender: male  \\
            Age: 35-44  \\
            Occupation: other  \\
            Location: Cincinnati, OH  \\
        
        \# \textbf{History}  \\
            A movie viewed just now,
            Pink Floyd - The Wall (1982) Drama|Musical|War, 
            My Behavior: Rating: 3.0/5.0. \\
            A movie viewed just now,  
            My Bodyguard (1980) Drama,
            My Behavior: Rating: 3.0/5.0.  \\
            A movie viewed just now,
            Agnes of God (1985) Drama|Mystery, 
            My Behavior: Rating: 3.0/5.0.  \\
            A movie viewed just now,
            Color of Money, The (1986) Drama, 
            My Behavior: Rating: 4.0/5.0.  \\
            A movie viewed just now,
            Betrayed (1988) Drama|Thriller, 
            My Behavior: Rating: 3.0/5.0.  \\
            A movie viewed just now,
            Radio Days (1987) Comedy|Drama, 
            My Behavior: Rating: 4.0/5.0.  \\
            A movie viewed just now,
            Mosquito Coast, The (1986) Drama, 
            My Behavior: Rating: 3.0/5.0.  \\
            A movie viewed just now,
            Brother from Another Planet, The (1984) Drama|Sci-Fi, 
            My Behavior: Rating: 4.0/5.0.  \\
        
    \# \textbf{Exposure List}  \\
    A. Psycho (1998) - Crime|Horror|Thriller  \\
    B. Footloose (1984) - Drama   \\
    C. Resurrection Man (1998) - Drama|Thriller  \\
    D. Curtis's Charm (1995) - Comedy|Drama  \\
    E. Psycho II (1983) - Horror|Thriller  \\
    \end{tcolorbox}
    
The model's output consists of the simulated user's decision-making process and the final selected option.

\begin{tcolorbox}[colback=myblue!5!white,colframe=myblue!95!white, fontupper=\slshape, title=Example of Output]
\scriptsize
\textbf{Thought}: \\
\textbf{Stimulus}: I'm feeling a bit restless and in the mood for a movie to unwind after a long day. I'm in the mood for something that's not too heavy but also not just another typical comedy or action film. \\
\textbf{Stimulus Factors}: Emotional State, Time of Day \\

\textbf{Knowledge}: From the exposure list, I noticed that "Footloose" (1984) and "Resurrection Man" (1998) are dramas, which aligns well with my recent movie preferences. "Footloose" seems to have a strong narrative and a mix of drama and possibly some action, which I find appealing. "Resurrection Man" is a thriller and also a drama, which also fits my current mood. \\
\textbf{Knowledge Factors}: Product/Service Attributes, Information Source \& Presentation \\

\textbf{Evaluation}: I think "Footloose" is the better choice because it seems to have a good balance of drama and a touch of action, which I think will be a nice break from the heavy dramas I've been watching. The action elements in "Resurrection Man" are intriguing, but I think "Footloose" will be more in line with my recent viewing history and preferences. \\
\textbf{Evaluation Style}: Logical \\

\textbf{Behavior: B}
\end{tcolorbox}
    
    
The results demonstrate that the fine-tuned LLM is capable of analyzing the recommendation task from both the \textit{stimulus} and \textit{knowledge} perspectives, thereby exhibiting a human-understandable reasoning process. 
Ultimately, it evaluates the available options based on logical reasoning and arrives at the most accurate selection.

In Appendix \ref{Append:Factors}, we conduct a statistical analysis of the factors referenced by the model when processing recommendation tasks. Here we provide serveral additional examples of the decision-making process of the user simulator.

\begin{tcolorbox}[colback=myblue!5!white,colframe=myblue!95!white, fontupper=\slshape, title=Example of Decision-Making Process]
    \scriptsize
    \textbf{Thought}: \\
    \textbf{Stimulus}: I am feeling a bit restless and in need of some entertainment after a long day at work. I am in the mood for something that resonates with my past viewing preferences and has a good mix of drama and comedy. \\
    \textbf{Stimulus Factors}: Emotional State, Past Experience \\
    
    \textbf{Knowledge}: From the exposure list, I have considered a few options. "American Graffiti" stands out as it combines drama and comedy, genres that I have rated highly in the past. "Outlaw Josey Wales" is another interesting option, but the Western genre is not as appealing to me as the dramas or comedies. "Blood Feast" is a horror film, which is not my preferred genre. "Little Big League" is a children's movie, which is not aligned with my current mood. \\
    \textbf{Knowledge Factors}: Personal Relevance, Past Experience, User Preferences/History \\
    
    \textbf{Evaluation}: I am leaning towards "American Graffiti" because it offers a blend of drama and comedy, genres that I have found to be more engaging and relatable. The other options do not align as closely with my current mood or past viewing history. \\
    \textbf{Evaluation Style}: Logical \\
    
    \end{tcolorbox}

\begin{tcolorbox}[colback=myblue!5!white,colframe=myblue!95!white, fontupper=\slshape, title=Example of Decision-Making Process]
    \scriptsize
    \textbf{Thought}: \\
    \textbf{Stimulus}: I'm feeling a bit nostalgic and in the mood to discover some new music that aligns with my taste in alternative and indie rock. I'm also influenced by the social aspect of seeing artists my friends have listened to. \\
    \textbf{Stimulus Factors}: Emotional State, Social Factors \\
    
    \textbf{Knowledge}: From the exposure list, I noticed that Belle and Sebastian have tags that resonate with my interest in alternative and indie rock, while The Microphones also caught my attention with their post-rock and rock tags. I also noted that one of my friends has listened to The Microphones, which adds to the appeal. \\
    \textbf{Knowledge Factors}: Product/Service Attributes (Genre, Social Proof), Information Source \& Presentation (Tags, Social Factors) \\
    
    \textbf{Evaluation}: I feel that Belle and Sebastian have a sound that aligns well with my taste in alternative rock, and the fact that a friend has also listened to The Microphones makes me curious to explore their music as well. \\
    \textbf{Evaluation Style}: Logical \\
    
    \end{tcolorbox}


\end{document}